\documentstyle[preprint,tighten,aps]{revtex}    
\begin{document}           %
\draft
\preprint{\vbox{\noindent
To appear in Astroparticle Physics \hfill INFNFE-10-95\\
          \null\hfill  INFNCA-TH9512}}
\title{Just so? \\
Vacuum oscillations and  MSW: an update.
      }
\author{
         E.~Calabresu$^{(1,2)}$,
         N.~Ferrari$^{(3,4)}$,
         G.~Fiorentini$^{(5,6)}$
         and M.~Lissia$^{(2,1)}$
       }
\address{
$^{(1)}$Dipartimento di Fisica dell'Universit\`a di Cagliari, I-09124
        Cagliari, Italy\\
$^{(2)}$Istituto Nazionale di Fisica Nucleare, Sezione
        di Cagliari, I-09128 Cagliari, Italy\\
$^{(3)}$Dipartimento di Fisica dell'Universit\`a di Milano, I-20133 Milano\\
$^{(4)}$Istituto Nazionale di Fisica Nucleare, Sezione di Milano,
        I-20133 Milano \\
$^{(5)}$Dipartimento di Fisica dell'Universit\`a di Ferrara, I-44100 Ferrara \\
$^{(6)}$Istituto Nazionale di Fisica Nucleare, Sezione di Ferrara,
        I-44100 Ferrara
        }
\date{June 1995}
\maketitle                 
\begin{abstract}
We find that vacuum oscillations (VO), large-mixing-angle and
small-mixing-angle MSW solutions to the solar neutrino problem (SNP)
give all very good fits to the most recent results.
Measurements of the $^7$Be flux can, in some cases, discriminate between
different solutions to the SNP; in particular, VO allow
$^7$Be fluxes almost as large as the one predicted by the SSM.
We find that no evidence for seasonal variations can be extracted from
present data, but that the large statistics of SuperKamiokande should
make possible to study a
significant portion of the presently allowed parameter space by just
looking for seasonal variations.
We also discuss the Borexino potential for detecting seasonal variations,
which looks really impressive.
\end{abstract}
\narrowtext

\section{Introduction}
\label{intro}
Since the first results from the radiochemical chlorine
experiment of Davis and his collaborators, the considerable deficit in
the solar neutrino flux respect to the corresponding theoretical predictions
has been the essence of the  solar neutrino problem (SNP).
Presently, there are four running
experiments~\cite{Davis,Kamio,GALLEX,SAGE},
which measure
different contributions to the solar neutrino flux, and all four of
them report rates that are just a fraction of the corresponding
theoretical  prediction.
Several
analysis~\cite{Cast93b,Bere93,Cast94a,Hata1,Bere94a,Bahcall94,%
Smirnov,Hata2,Degli94},
which use the information from all the
experiments and only mild theoretical assumptions, show that the SNP is
now practically independent of the details
of the solar models and strongly suggest a modification of the neutrino
properties.

Giving neutrinos a small flavor-non-diagonal mass offers the
attractive possibility of reconciling theory and experiments
with a minimal change of the standard neutrino physics. In the context
of a two flavor analysis there exist two kind of solutions:
vacuum oscillations (VO), also known as just-so oscillations, and
Mikheyev-Smirnov-Wolfenstein (MSW) matter enhanced oscillations (see
Refs.~\cite{Bahcall89,Bahcall95a,Haxton95} for a few reviews
and lists of references).

In this paper we want to consider the present status of both kind of
solutions, VO and MSW, in the light of the latest experimental
results~\cite{Davis,Kamio,GALLEX,SAGE}.
We also consider the possibility of VO into sterile neutrinos.
We devote particular care to the
solution of the neutrino evolution, which we perform numerically for
masses both in the MSW and in the VO range with the methods described in
Ref.~\cite{Fiore}, and to the statistical analysis of the data, especially
concerning their possible time dependence. On the other hand, we do not
consider energy spectrum deformations: discussions on this point can be
found in Refs.~\cite{Hata94,Rossi94}.

The main differences between
recent similar analyses~\cite{Petcov94,Hata94,Rossi94}
and ours are the use of the new GALLEX data~\cite{GALLEX} and the
use of the recalibrated chlorine data~\cite{Davis},
whose average is now a standard deviation higher than the value that has been
quoted in the past few years.

Nowadays experiments explicitly aimed at measuring the flux of $^7$Be
neutrinos are a recognized
priority~\cite{LNGS1,LNGS2}: we consider the impact of
such measurements, and in particular of Borexino~\cite{Arpe92}.

More specifically, in this paper we want to address the following
questions:\\
(1) What are the chances of VO and MSW in the light of the data
    presently available?\\
(2) Is it possible that future experiments, and in particular Borexino,
    will be able to measure a large
    (close to the SSM prediction) $^7$Be flux?\\
(3) Can a measurement of the $^7$Be line discriminate between VO and MSW
    solutions within the regions of parameters that are allowed by present
    data?\\
(4) Can a measurement of the pep line be of significant help in this
    respect?\\
(5) It is well-known that seasonal modulations of the expected signal occur
    naturally in the VO scenario: is there any positive hint of such
    modulations in the available data?\\
(6) How much should the statistics of running experiments be improved in order
    to have a significant sensitivity to seasonal modulations?\\
(7) What region of the parameter space can be explored by Borexino just by
    searching for time modulations? Does this region include the parameters
    that are favored by present data?

The paper is organized as follows. In Sec.~\ref{sec2}
we discuss vacuum oscillations
in the light of present results, both the expected signals averaged over the
year and their seasonal variations.
Section~\ref{sec3} is dedicated to the beryllium and pep signals:
the yearly averaged and the time modulations are considered
with emphasis placed on the upcoming Borexino experiment signal.
In Sec.~\ref{sec4} we compare the VO and MSW solutions to the SNP. Finally,
we summarize our conclusions in Sec.~\ref{concls}. In the Appendix we collect
those details and tests of our calculation that would hinder the main flow
of the presentation.

\section{Vacuum oscillations in the light of the latest results}
\label{sec2}
Our predictions have been derived in the context of standard solar models
(SSM) and vacuum mixing between two species of neutrinos.
Specifically, we use as a reference model the
SSM of Bahcall and Pinsonneault~\cite{Bahcall92} (BP92). We have also made
several tests with the most recent model by Bahcall and
Pinsonneault~\cite{Bahcall95b} (BP95), which includes diffusion of
heavy elements, and we have found that our conclusions would not be
significantly modified. We consider both
the case when the electron neutrino oscillates into an active neutrino,
which then contributes to the Kamiokande signal, and the case when
the electron neutrino oscillates into a sterile neutrino.

It is well-known that there are two ranges of masses for which two-flavor
neutrino mixing has the possibility of explaining present data:
the MSW mechanism,  which we consider in Sec.~\ref{sec4}, works for masses
$ 0.1 (m\text{eV})^2\le \Delta m^2\le 500 (m\text{eV})^2$,
while the so-called VO or just-so solution, which is the topic of this section,
needs masses $ 30(\mu\text{eV})^2\le \Delta m^2\le 200 (\mu\text{eV})^2$.

As mentioned in the introduction, we evolve neutrino amplitudes numerically
from the production to the detection point with a hamiltonian that explicitly
contains the interaction with matter. For a given choice of mixing parameters
$(\sin^2 2\theta,\Delta m^2)$, the result is a survival probability
$P_{\nu_{\text{e}}\to \nu_{\text{e}}} (E, \bbox{r}, t)$ that depends on the
neutrino energy $E$, the production point $\bbox{r}$ and the time of the year
$t$ (it actually depends only on the distance between the production point
and the position of the detector at a given time of the year).
Incidentally, we verified that the simpler
procedure of evolving neutrino amplitudes in vacuum over the distance from
the surface of the Sun to the detector gives results within a few percents of
the exact evolution, i.e. matter has the only effect of
inhibiting the oscillations inside the Sun in the mass range of VO.

Let us define
the cross sections of the $i$-th experiment for detection of electron
neutrinos $\sigma^{\text{e}}_i(E)$ and for detection of the
other-flavor neutrinos $\sigma^{\mu}_i(E)$;
of course, $\sigma^{\mu}_i(E)=0$ when the detection is only through
charged current (CC) or if electron neutrinos oscillate
into sterile neutrinos. In addition, let $\Phi_l(E,\bbox{r})$ be the
differential neutrino flux per unit of energy and unit of volume
produced by the $l$-th reaction ($l=$ pp, pep, $^8$B, $^7$Be,
$^{13}$N and $^{15}$O) with energy $E$ at point $\bbox{r}$.
Then we calculate the expected signal in the $i$-th detector at time $t$
\begin{mathletters}
\label{thsgnl2}
\begin{equation}
S^{\text{th}}_i(t) =
\sum_{l} \int_{V} d^3 \bbox{r}
\int_0^{\infty} \!dE\, \Phi_l (E,\bbox{r})
 \left[ \left (
                \sigma^{\text{e}}_i(E) - \sigma^{\mu}_i(E)
        \right) P_{\nu_{\text{e}}\to \nu_{\text{e}}} (E, \bbox{r}, t)
        + \sigma^{\mu}_i(E)
\right]\, ,
\label{thsgnlt}
\end{equation}
and its average over the whole year ($T=$ one year)
\begin{equation}
\overline{S}^{\text{th}}_i \equiv
\langle S^{\text{th}}_i \rangle_{T}=
\frac{1}{T} \int_0^{T} \! dt\, S^{\text{th}}_i(t) \, .
\label{thsgnla}
\end{equation}
\end{mathletters}
\subsection{The yearly averaged information}
In this subsection we compare theoretical predictions for the signals in
the detectors averaged over the year,
$\overline{S}^{\text{th}}_i $,
with the corresponding experimental rates from the
four solar neutrino experiments (see Table~\ref{tbl1}).

We use the following likelihood formula to compare theoretical predictions
and experimental data:
\begin{equation}
\chi^2 = \sum_{i=1}^{N_{\text{exp}}}
     \left(\frac{S^{\text{exp}}_{i} - \overline{S}^{\text{th}}_{i}}
           {\Delta S^{\text{exp}}_{i}}\right)^2 \quad ,
\end{equation}
where $S^{\text{exp}}_{i}$ is the experimental signal for the $i$-th
experiment with its corresponding total error $\Delta S^{\text{exp}}_{i}$
(statistical and systematic errors have been
quadratically combined). This $\chi^2$ is a function of the parameters
$(\sin^2 2\theta, \Delta m^2)$ through the theoretically predicted signal
$S^{\text{th}}_{i}$ that has been defined in Eqs.~(\ref{thsgnl2}).
Then we define the regions allowed at the $x$\% C.L. by
$\chi^2-\chi^2_{\text{min}}<
\chi^2_{\text{crit}}(x)$, where $\chi^2_{\text{min}}$ is the
minimum value of $\chi^2$ as the two parameters are varied, and
$\chi^2_{\text{crit}}(x)$ is the critical value for 2 degrees of freedom
(for any $x$, there is a well defined $\chi^2_{\text{crit}}(x)$ such that
one has a $x$\% probability that statistical fluctuations give
a $\chi^2 > \chi^2_{\text{crit}}(x)$).

We find $\chi^2_{\text{min}}/d.o.f.=2.4/2$ and
$\chi^2_{\text{min}}/d.o.f.=6.2/2$ for active and sterile neutrinos,
respectively.

All in all,
a mechanism where the electron neutrino mixes with an active
neutrino provides a very good description of the most recent solar
neutrino data, while the mixing with a sterile neutrino is excluded
at the 95\% C.L..

In Figs.~\ref{fig1} we show the region of parameters that the combination
of the 4 experiments allows at the 68\%
($\chi^2\le \chi^2_{\text{min}} + 2.28$) and 95\%
($\chi^2\le \chi^2_{\text{min}} + 5.99$) C.L. for oscillations
into active and sterile neutrinos. We have used fluxes from the SSM of BP92
with no theoretical errors. In the same figures we also show the points
of best fit.

In Figs.~\ref{fig2} we show the allowed area by each of the 4 experiments
individually; here the regions are defined by
$\chi^2 < 0.99 $  and $\chi^2 < 3.84 $ at the 68\% and 95\% C.L.,
respectively.
For the Kamiokande experiment the  cases of mixing with active
(68\% and 95\% C.L. regions) and with
sterile (only  95\% C.L. region) neutrino are shown on the same figure (c).

The comparison between the different areas of Fig.~\ref{fig2} is
interesting: gallium experiments allow a wide region of parameters, and
the best fit
is mostly the result of compromising the higher suppression implied by
the chlorine datum and the lower one needed by the Kamiokande datum.
The combination of these  constraints roughly selects parameter regions
where the contribution to the chlorine signal
from the neutrino spectrum below about 5~MeV (beryllium, CNO and part
of the boron flux)
is more suppressed than the contribution from the boron flux above 5~MeV.
If the electron neutrino mixes with a sterile neutrino,
the boron flux above 5~MeV needs to be suppressed even less to reproduce
the Kamiokande data, the contrast with the chlorine data increases and
the fit becomes worse.

So far we have discussed the best fit point; the study of a few other solutions
with reasonable confidence level is also of interest. To get an idea of the
signal to be expected in future $^7$Be experiments, we show in
Table~\ref{tbl2}, together with information about the best fit points, also
the points corresponding to minimum and maximum yearly averaged $^7$Be
signal within the 90\% C.L. region in the third and fourth column.
Contrary to general expectation, we see that the chance of a $^7$Be
signal close to the SSM prediction is not at all excluded.
On the other hand, a $^7$Be flux smaller than about 1/4 of the SSM prediction
would be a strong indication against VO; in fact
the $^7$Be signal must be larger than about 2/5 of SSM prediction
if neutrinos are active and detection is by charged current (CC) plus
neutral current (NC) as in Kamiokande or Borexino.

Our fits to the experimental data are somewhat better than the ones
shown in the recent literature~\cite{Petcov94,Hata94,Rossi94}.
The only important differences between our calculation and the others
are in the gallium and chlorine data. For the gallium data, we use
the most recent value quoted by the GALLEX Collaboration~\cite{GALLEX},
and include in the analysis the SAGE value~\cite{SAGE}.
For the chlorine experiment we use the most recent
value~\cite{Davis}, which happens to be a standard deviation higher than it
used to be as a consequence of the recalibration of all chlorine data.
We verified that our use of a new and higher chlorine datum explains
our better fit, as expected by the fact that
the fit in this part of the Bethe plane is mostly constrained by the
chlorine and Kamiokande experiments.

The uncertainties of the SSM parameters imply corresponding
uncertainties in the prediction of the neutrino fluxes and, therefore, in
the theoretical signals. Since the same parameter can affect more than one
flux, and the same flux can contribute to more than one signal, the resulting
errors are strongly correlated.
These correlations have been taken into account in two ways: (a) by using
the full correlation matrix (for details see Ref.~\cite{Cast94b})
and (b) by making a Monte Carlo simulation (see Appendix).
The two approaches give essentially the same result, and in
Fig.~\ref{fig3} we present only the iso-curves obtained with the first method.

With respect to the case without theoretical errors, the position of the best
fit is the same, and the quality of the fit is basically unchanged:
$\chi^2_{\text{min}}$ is now 2.2 against the previous value of 2.4,
for active neutrinos, while $\chi^2_{\text{min}}$ is 6.1 against 6.2,
for sterile neutrinos.

Physically, the quality of the fit does not improve significantly since
at the point of best fit the best compromise has already been reached, and
a variation of any of the major fluxes in either direction only makes
the fit worse.
Taking into account the theoretical uncertainties makes only
the confidence regions larger, as it is shown in Fig.~\ref{fig3}. In fact,
contrarily to what happens for the best fit point, the extra freedom
in choosing the fluxes allowed by the theoretical uncertainties can
help to improve the fit in those points where by definition the balance
between the fluxes is not optimal.

To test the stability of the VO solution, we have also determined the best fit
solutions when the temperature $T$ or the astrophysical factor $S_{17}$ are
varied with respect to the values of the SSM. We report results only
for the case of active neutrinos in Table~\ref{tbl3}; they indicate that:\\
(1) if $T$ varies by $\pm 2\%$ we still find reasonable fits;\\
(2) increasing the temperature is preferred over decreasing it;\\
(3) increasing $S_{17}$ by 50\%
    improves the quality of the fit
    (here and in the previous case
    a higher boron flux in the Solar Model
    makes the suppression needed by Kamiokande closer to the
    one needed by the chlorine experiment);\\
(4) decreasing $S_{17}$ by 50\% (i.e a lower boron flux in the Solar Model)
    makes the fit worse;\\
(5) except for the last case, the position of the best fit does not change
    much, consistently with the fact that we have found the same best fit
    point with and without theoretical uncertainties (consider that
    in this test we have made variations larger than the ones allowed
    by the uncertainties).\\
Similar stability tests in the case of the MSW solution can be found in
Ref.~\cite{Bere94b}.
\subsection{Any signal of seasonal variations?}
It is well-known that VO predict that the survival probability
should change during the year. In fact, in the range of masses of interest:
$ 30 (\mu \text{eV})^2\le \Delta m^2\le 200 (\mu \text{eV})^2$,
the oscillation wavelength ( in astronomical units $L= 16.53 E/\Delta m^2$,
where the energy $E$ is in MeV and the mass difference $\Delta m^2$
in $(\mu$eV$)^2$) is comparable with the seasonal change of the
distance between the Sun and the Earth (about 4\% of its average value).

The detection of this expected seasonal variation is extremely important
to test the viability of the VO mechanism. In this subsection we
want to study whether any temporal information can already
be extracted from the present data.

To simplify the treatment of the data and improve their graphical impact,
we have divided the year into six bins as shown in Fig.~\ref{fig4}.
At any rate, since all
runs are at least one-month long, we find no significant statistical gain
in using the information of the individual unbinned runs.
Clearly, we face at least two problems:
(1) To which bin(s) do we assign runs whose duration overlaps with more
than one bin?
and (2) How do we combine (``average'') individual runs that have been
assigned to a bin and, therefore, define a rate for the bin?
Different procedures, which we briefly describe in the Appendix,
have been tested with no significant differences.
In the following, we shall consider the second approach described in
the Appendix.
Data have been taken from Refs.~\cite{Davis2,Kamio2,GALLEX2,SAGE2}.

In Figs.~\ref{fig5} we compare these binned data with the
expected variation of
the signal for several typical values of the mixing parameters. The variation
due to the geometric factor, which is known exactly, is not included in the
theoretical calculations and has been taken out from the experimental data,
since we are interested only in the time dependence caused by the oscillation.
We have also renormalized data to the yearly averaged signal:
we want to separate  the effect of the time dependence from the change in
the average.
It is already clear from these figures that statistical
fluctuations are at the moment much larger than the expected variations.

Let us make more quantitative the statement about the present chance
of detecting any temporal variation.
Let us define
\begin{equation}
S^{\text{th}}_i(j) = \frac{1}{T_{\text{bin}}} \int_{\text{bin}(j)} \!\! dt
\, S^{\text{th}}_i(t) \quad ,
\label{thsgnlb}
\end{equation}
where the integral is over the $j$-th bin, as defined in Fig.~\ref{fig4},
$T_{\text{bin}}$ is the length of each of the bins (1/6 of year, since
the year has been uniformly divided among the bins) and
$S^{\text{th}}_i(t)$ had already been defined in Eq.~(\ref{thsgnlt}). We also
need
$\overline{S}^{\text{th}}_i=\frac{1}{6}\sum_{j=1}^6 S^{\text{th}}_i(j)$,
which has already been defined in Eq.~(\ref{thsgnla}).
Similarly,
$S^{\text{exp}}_i(j)$ and $\Delta S^{\text{exp}}_i(j)$ are
the value and error of the $j$-th bin for the $i$-th experiment.

Then we calculate $\chi^2_a$ by using the average theoretical signal:
\begin{mathletters}
\begin{equation}
\chi_a^2 = \sum_{i=1}^{N_{\text{exp}}} \sum_{j=1}^{6}
     \left(
    \frac{S^{\text{exp}}_i(j) - \overline{S}^{\text{th}}_i}{\Delta
      S^{\text{exp}}_i(j)}
    \right)^2 \quad ,
\label{chia}
\end{equation}
while an alternative $\chi^2_b$ is calculated by using the binned theoretical
signals:
\begin{equation}
\chi_b^2 = \sum_{i=1}^{N_{\text{exp}}} \sum_{j=1}^{6}
     \left(
    \frac{S^{\text{exp}}_i(j) - S^{\text{th}}_i(j)}
          {\Delta S^{\text{exp}}_i(j)}
    \right)^2 \quad .
\label{chib}
\end{equation}
\end{mathletters}
If the experimental signals have the time dependence predicted by the VO
mechanism, and the statistics were large enough,
the minimum of $\chi^2_b$ should be significantly lower than the minimum
of $\chi^2_a$. Note that both definitions treat
the experimental data exactly on the same footing, so that we can safely
ascribe any difference to the time variation.

There is no significant difference between the two $\chi^2$'s, which is not
surprising after Figs.~\ref{fig5}.
At the point of best fit we find $\chi^2_a/d.o.f= 10.5/22$ (constant fit),
and $\chi^2_b/d.o.f= 11.4/22$ (time dependent fit).
The 68\% and 95\% confidence regions
remain also basically unchanged, as it is shown in Figs.~\ref{fig6}.
We have performed the same exercise also for sterile neutrinos:
constant and time dependent theoretical predictions produce basically the
same confidence regions, which are also similar to those shown
in Fig.~\ref{fig1}, therefore we do not show them again.

Different definitions of the confidence regions (e.g. via Monte Carlo
simulations) do not change our conclusion; for more details see Appendix.

After having established that the statistics of present experiments is
insufficient to detect seasonal variations, let us consider what region of
the parameter space could be explored if this same statistics were improved
by a factor $x$. The formula we use to make this estimate is also
reported in the Appendix, i.e. Eq.~(\ref{nest}).\\
(a) \underline{Gallium experiments}:
 We take as a reference ($x=1$) the statistics
presently collected
by GALLEX (30 tons $\times$ 3 years). For $x=10$, (e.g. GALLEX + SAGE
running for a total of about 12 years)
only a very small region of the Bethe-plane,
which does not even contain the best fit point,
could be investigated (see Fig.~\ref{fig7}a).
On the other hand, an improvement factor $x=25$ would allow to explore most
of the presently favored region (see Fig.~\ref{fig7}b).\\
(b) \underline{Kamiokande}: We present for active neutrinos in Fig.8 the cases
$x=100$ and $x=170$. Note that in the last case one can explore the best fit
proximity. For sterile neutrinos, $x=100$ would suffice to reach the best
point.\\
Superkamiokande, which is expected to collect about 100 times as many events
per day as Kamiokande, should thus reach a significant sensitivity. In
addition,
with such large statistics, it will be possible to study spectral deformation
{}~\cite{Rossi94}.\\
(c) \underline{Homestake}:
The chlorine experiment would need $x=10$ to be sensitive to
the seasonal variations in the region of the best fit point.

\section{Beryllium (and pep) neutrinos in VO}
\label{sec3}
The signal from solar neutrinos produced in the reaction
$^7\text{Be}+\text{e}^-\to ^7\text{Li}+\nu_{\text{e}}$ is of particular
interest in the study of neutrino oscillations. One of the reasons is
that this signal has a
fixed energy (we consider only the more intense line at 0.861~MeV)
making more direct the analysis in terms of the
oscillation parameters; in particular, its dependence on
the Sun-Earth distance is not smeared out by integrating over the energy
spectrum.

In this section we analyze the prediction of the vacuum oscillation mechanism
for the $^7$Be neutrino flux and we also comment on the significance of a
pep line intensity measurement.
\subsection{The expected yearly averaged beryllium and pep signals}
In Fig.~\ref{fig9} we present the yearly averaged suppression for
$^7$Be neutrinos superimposed with the 95\% C.L. regions, taken from
Fig.~\ref{fig1}.

One sees that VO can accomodate all experimental data without requiring
drastic reduction of beryllium flux. In other words, the $^7$Be signal in
experiments is not necessarily very low, if VO are the right explanation.

In Table~\ref{tbl2} we present the minimum and maximum expected
signal (normalized to the SSM prediction) both for the case of
CC detection and CC + NC detection.
One sees that signals up to almost the SSM prediction are possible.
On the other hand, a strongly suppressed $^7$Be signal
(below about 0.2 for CC detection or below about 0.4 for
CC + NC detection) would be a clear indication against VO.

A measurement of the pep signal would be extremely interesting for three
reasons:
 (1) the production of pep is very weakly dependent on solar models;
 (2) pep neutrinos are, similarly to the beryllium neutrinos, monochromatic
     and therefore very sensitive to oscillation mechanisms;
 (3) differently from the $^7$Be, $^8$B or pp fluxes,  there is essentially
     no experimental information available about the pep flux so far.

Analogously to Fig.~\ref{fig9} for the beryllium flux, in Fig.~\ref{fig10}
we present the iso-suppression curves for the pep flux.
Again we find that the pep flux can take essentially
any value, and VO are still consistent with experimental data.
The best fit points correspond to a strong suppression of
pep (see Fig.~\ref{fig11}).
However, there is no strict correlation between survival
probabilities of the $^7$Be and pep neutrinos, as can be
seen from Fig.~\ref{fig11}:
in fact in this range of parameters the survival probability oscillates
rapidly as a function of the neutrino energy (compare with the
different situation for the MSW solution).

\subsection{Seasonal modulation and future experiments}
The specific characteristic of a solution of the SNP by
means of VO
is the expected seasonal variation. This variation
should be particularly pronounced for the monoenergetic beryllium signal.
A detection of
this variation by future experiment would provide strong evidence for this
mechanism.\\
Seasonal variations of the pep signal would also be clearly interesting, but we
feel that their detection is beyond the experimental capability, at least
in the foreseeable future, and we do not discuss them.

In Fig.~\ref{fig12} we show the expected variations for a few representative
choices of the parameters within the 90\% C.L. allowed regions:
the best fit case of maximal (yearly averaged)
beryllium signal and the point of maximal modulation of the beryllium signal.
Clearly the modulation is less strong if both CC and NC are detected
(Fig.~\ref{fig12}a). The variation due to the geometric
effect (which can be eliminated exactly from the data)
is not included to make clear the size of the relevant physical
effect.

The modulation (defined as the difference between the signal in the
winter, i.e. bins 1,2 and 3 in Fig.~\ref{fig4}, and in the summer semester) is
of the order of 30\% of the average value at the best fit parameters.
The maximal variation is not much larger indicating that this size of
modulation is typical.

On the other hand, the curve corresponding to the parameters that give
the maximal averaged value of the beryllium signal shows almost no modulation:
this gives us a warning that VO do not necessarily imply a significant
modulation (the obvious explanation being that the maximum of the
survival probability is a stationary point, i.e. there is no
modulation at the first order).

The possibility of seeing the seasonal variation of the signal depends on the
magnitude of the effect, on the overall statistics and on the
signal-to-background ratio. The formula we use to make our estimates
is reported in the Appendix, Eq.~(\ref{n0est}).

In Fig.~\ref{fig13} we present the region that can be explored at the 3 sigma
level by Borexino assuming 50 events per day for standard neutrinos and
1000 running days ($N_0=50000$). We consider two cases for the signal to
background ratio:

\noindent
(a) $s_0 = 5$, corresponding to about 11 background events per day
(high purity case~\cite{Arpe92} ).

\noindent
(b) $s_0 = 5/3$, i.e. a three time worse background~\cite{bonetti}.

Results are shown for active neutrinos (we recall that for sterile neutrinos
the modulation is larger). We conclude that Borexino covers very well the
interesting region. This conclusion is true also in the (pessimistic) case
(b) .

\section{Vacuum oscillations, MSW and what else?}
\label{sec4}
Several mechanisms have been proposed to resolve the SNP. It is natural to
compare the vacuum oscillation mechanism with at least the other most
popular one: the MSW solution.

The MSW mechanism is very robust. It always provides a very good fit to
the experimental data, and the solutions are relatively stable against changes
in the SSM. Since the experimental data have changed somewhat,
we thought better to recalculate the best fit and confidence regions
also for the MSW mechanism.
The details of the calculations are the same as in
Ref.~\cite{Fiore}. The theoretical errors have been
introduced taking into proper account their correlation using the same
procedure as in Ref.~\cite{Cast94b}.
Results are shown in Fig.~\ref{fig14} with and without theoretical errors.
In Table~\ref{tbl4} we report
information about the MSW solution analogous to the one reported in
Table~\ref{tbl2} for the VO solution. We only consider active neutrinos.

The small-mixing-angle solution (best fit) remains more or less unaffected by
the new data. It is the large-mixing-angle solution that changes most,
due mainly to the one-standard-deviation increase of the chlorine
data: this solution now appears already at the 74\% C.L.
and has a 95\% C.L. region larger than before.

The MSW best fit ($\chi^{2}/d.o.f. = 0.5/2$) looks better in respect
to the VO best fit ($\chi^{2}/d.o.f. = 2.4/2$); nevertheless the
two fits are both very good. In perspective,
there are several features that could allow to distinguish between
the two solutions.
The MSW  solution gives a signal constant in time, while the VO one
gives a signal that, in general, depends on the time of the year as we
just discussed.
Moreover, the deformation of the energy spectrum is different in the two
cases~\cite{Hata94}.
We concentrate here on what can be learned from yearly averaged measurements
of the intensity of the Be line and, possibly, of the pep line as well.

In Fig.~\ref{fig15} we present the expected range of suppression
(whithin the 90\% C.L. region)
of the beryllium signal in a CC + NC detector, e.g. Borexino, in the case
in which neutrinos mix with an active flavor (VO, large and small mixing
angle MSW) or are standard.

First of all, it is worth observing
that high values (close to SSM) of the beryllium signal are foreseeable.
In addition:\\
(1) a $^7$Be signal larger than about 0.7 of the SSM prediction would
    only be consistent with the VO solution;\\
(2) in the intermediate range,
    $0.5 < \Phi_{\text{Be}}/ \Phi_{\text{Be}}^{\text{SSM}} < 0.7$,
    both VO and large angle MSW solutions are acceptable; \\
(3) in the range above about 0.2 and below about 0.4 basically only
    the small angle MSW solution is consistent with the data;\\
(4) if the $^7$Be signal is less than about 0.2 of the SSM prediction, the
    solution to the SNP cannot be mixing with an active neutrino;\\
(5) if the signal is below about 0.14, there is also a small
    chance of a solution with standard neutrinos; this bound comes from
    the combination of all four experiments and the luminosity sum rule by
    using methods similar to those in Refs.~\cite{Degli94,Hampel1,Hampel2}.\\
In case of a $^7$Be measurement that cannot discriminate between VO and MSW,
additional information can clearly be derived from the temporal dependence of
the signal, which is particularly strong when VO give suppressions around
0.5. Furthermore, a measurement of the pep flux could be illuminating. In fact,
the MSW mechanism, differently from VO, implies a strong correlation
between pep and beryllium signals, as it is clearly shown in Fig.~\ref{fig11}.

In Fig.~\ref{fig11} we show the predicted value of the pep vs. the
$^7$Be signal (in fractions of their SSM value)
in the case of VO (crosses) for
parameters inside the 90\% confidence region assuming a CC+NC detector
and active neutrinos.
We compare them with the values predicted by other possible solutions to the
SNP: astrophysical (non standard solar models) and MSW (small and large
mixing angle). We find that VO are compatible with almost any suppression
of the pep flux, as it was the case with the $^7$Be flux.
Nevertheless, the combination of the $^7$Be and pep flux information
tell us that vanishing $^7$Be and pep fluxes characterizes the
small-mixing-angle MSW solution, while an equal suppression of the two fluxes
points towards the large-mixing-angle MSW solution. Any other combination
appears to favor the VO solution.
\section{Conclusions}
\label{concls}

We summarize here the basic conclusions of our discussion.

\noindent (1)
  In the light of the most recent experimental results both the VO solution to
the SNP and the large-mixing-angle MSW solution have been resurrected. In fact
both solutions are now good, for active neutrinos.

\noindent (2)
  We find that the $^7$Be signal predicted by VO can be quite large (almost
equal to the SSM prediction) within the region of parameters allowed by
present experiments.

\noindent (3)
 Measurements of the beryllium flux can, in several instances, clearly
 discriminate between different solutions of the SNP, see Fig.~\ref{fig15}.

\noindent (4)
 Measurements of the pep line could help and discriminate between
 cases for which knowledge of the only $^7$Be flux is open to more than
 one interpretation.

\noindent (5)
 We find that present data do not show any sign of seasonal variations.

\noindent (6)
 On the other hand, the high statistics of SuperKamiokande should allow a
 study of a significant portion of the parameter space including most
 of the region favored by present data, see Fig.~\ref{fig8}.

\noindent (7)
 Borexino potential of detecting seasonal variations is really exciting:
 this experiment should definitely either detect seasonal variations or
 essentially rule out the VO mechanism.
\acknowledgments
This work has been inspired by several discussions with R.~Barbieri,
E.~Bellotti and Z.~Berezhiani, which we gratefully acknowledge.
We express our deepest appreciation to V.~Berezinsky for organizing
a very stimulating meeting on beryllium neutrinos at LNGS~\cite{LNGS1}.
We thank J.~N.~Bahcall for sending us Ref.~\cite{Bahcall95b}.
We thank A.~Suzuki for answering questions on the Kamiokande data and
sending us the thesis in Ref.~\cite{Kamio2}.
We also thank S.~Bonetti for very kindly providing us with information
about Borexino.
We thank the Gallex Collaboration for letting us use some of their data.
We also thank B.~Ricci for useful remarks.
\appendix
\section*{}
\label{appx}
\subsection{Confidence regions and theoretical errors}
In this work we have defined the confidence regions by
$\chi^2< \chi^2_{\text{min}}+  \chi^2_{\text{crit}}$ where
$\chi^2_{\text{crit}}$ is the critical $\chi^2$ value for 2 degrees of
freedom corresponding to the required confidence level (C.L.).

We have also tested the alternative definition
$\chi^2< \chi^2_{\text{crit}}$ where now $\chi^2_{\text{crit}}$ is the
critical $\chi^2$ value for the total number of degrees of freedom:
4 in the case of the yearly averaged case, and 24 (4 experiments times
6 bins) in the case of binned data. In this case confidence regions
from the the yearly averaged and binned data are different. But when
we correctly compare analogous definitions, Eqs.~(\ref{chia}) and
(\ref{chib}), our conclusion about the lack of evidence for seasonal variations
remains unchanged; the same is true for all other conclusions.

An other possibility of defining the confidence regions is by Monte
Carlo simulation. This procedure has the great advantage of being very
flexible: we can easily take into account, for instance,
theoretical errors without any need of explicitly introducing the
correlation matrix, and any other additional information
(e.g. efficiencies) can similarly be introduced.

We generate a sample of 1000 simulated experimental results
according to a probability distribution that takes into account
the experimental statistical and systematic errors. When we want to
include theoretical errors, we also generate a sample of theoretical
models by letting the input parameters fluctuate according to
their uncertainties and calculating the effect of changing the parameters
on the SSM by using known scaling rules~\cite{Cast94b}.
For each simulated experimental result (and theoretical model when
required) we compute the value of $\chi^2$ and compare it to the
value obtained from the real data; we define the C.L.
as the percentage of times that the simulated value is
smaller than the real one. We find results that are not significantly
different from the ones obtained using the $\chi^2$ definition
used in the main text.

\subsection{Binning the experimental data}
As mentioned in the main text, binning poses
at least two questions:
(1) How do we treat runs whose duration overlaps with more than one bin?
and (2) How do we calculate  the rate of a given bin from its individual
runs?

The simplest answer to the first question is that each run is assigned
that bin which contains the mean time of exposure,
defined~\cite{Bahcall89} as
\begin{equation}
t_{\text{mean}}=t_{\text{start}}+\tau \log [1/2+1/2
\exp(t_{\text{end}}/\tau-t_{\text{start}}/\tau)] \, ,
\label{tmean}
\end{equation}
where $t_{\text{start}}$ and $t_{\text{end}}$ are the starting and ending
time of the run, and $\tau$ is the mean-life of the final state
($\tau=\infty$ for electron scattering experiments).
This definition is such that we expect half of the detected events
to occur between $t_{\text{start}}$ and $t_{\text{mean}}$ and half
between $t_{\text{mean}}$ and $t_{\text{end}}$.

We have tried a more refined approach. When the exposure time of a given
run overlaps with more than one bin, this run contributes to all
the interested bins with a weight proportional to the number of the recorded
events  which are expected to come from the bin (this number depends on
the length of the overlap and on the mean decay time).

A more sophisticated procedure, which correctly uses the full experimental
information, consists in applying a maximum likelihood analysis to
the time information of all events (signal and background) assigned to
a given bin. The disadvantage of this method is that one needs access to the
complete event structure of a given experiment to be able to apply
this kind of analysis.
Since we had access only to GALLEX complete experimental information,
we have verified that, at least for
the GALLEX experiment, the two procedures agree within the errors.

\subsection{Statistical sensibility to seasonal variations}
In this part of the Appendix we provide the framework for quantitative
and simple estimate of the sensitivity of present and future experiments
to the time dependence of the neutrino signal, by discussing the
behavior of neutrino signals averaged over half year.

Let us consider the electron neutrino survival probability averaged over
a six month period centered around the perihelion
$\langle p \rangle_{w}$ and around the aphelion $\langle p \rangle_{s}$.
For each experiment, $W_{w,s}$ are also averaged over the accessible
energy interval, being weighted with the $\nu_e$
cross section. We also define
the corresponding averaged
interaction probabilities
$W_{w,s}$, which for a CC + NC detector (Kamiokande and  Borexino) and
active neutrinos are about:
\begin{equation}
W_{w,s}=1/5 + 4/5 \langle p \rangle_{w,s} \, ,
\end{equation}
while for CC detectors or sterile neutrinos are obviously:
\begin{equation}
W_{w,s}=\langle p \rangle_{w,s} \, .
\end{equation}
In order to characterize the size times running time of the experiment,
we consider the total number $N_0$ of neutrino events that are
expected for standard neutrinos. For sufficiently long running times,
the experiment will collect a number of true neutrino $N_t$ given by:
\begin{equation}
N_t=N_0 \frac{W_{w}+W_{s}}{2}\, .
\end{equation}
If $N_b$ is  the number of expected background along the running time,
a quality factor of the experiment is $s_0 = N_0/N_b$, i.e. the ratio
of expected neutrino events, for standard neutrino, to background counts.

By dividing the running time in winter and summer semesters, one expects
in each semester a number of counts (neutrinos plus background) given by:
\begin{equation}
N_{w,s}= \frac{N_0 W_{w,s} + N_b}{2} \, ,
\end{equation}
where we assume that the background mean rate is constant during the year.
Seasonal modulations can be detected at the $n$-sigma level if the
difference $N_w - N_s$ exceeds at the corresponding level the
statistical fluctuation of the total number of counts:
\begin{equation}
\left| N_w - N_s \right| \ge n\sqrt{N_w+N_s} \, .
\label{fluct}
\end{equation}
For the four running experiments, the accessible region can be determined
approximately from the quoted signals $S$ and statistical error $\Delta S$
and the expected signal for standard neutrinos $S_0$. In fact, if we use
\begin{eqnarray}
\frac{N_t}{N_0}&=&\frac{S}{S_0} \\
       \nonumber \\
\frac{\sqrt{N_w+N_s}}{N_t}&=&\frac{\Delta S}{S}
\end{eqnarray}
and Eq.~(\ref{fluct}), we find
\begin{equation}
\left| W_w - W_s \right| \ge \frac{2n \Delta S}{S_0}\, .
\end{equation}
The region of parameters determined by this condition should obviously be
intersected with the region such that $W_w + W_s$
is consistent with the reported signals.

If the statistics is increased by a factor $x$ within the same experimental
apparatus, the accessible region becomes
\begin{equation}
\left| W_w - W_s \right| \ge \frac{2n \Delta S}{S_0\sqrt{x}}\, .
\label{nest}
\end{equation}

A more general expression, which is particularly useful for future
experiments like Borexino, can be obtained from Eq.~(\ref{fluct})
by a straightforward substitution:
\begin{equation}
\left| W_w - W_s \right| \ge \frac{2n}{\sqrt{N_0}}
                                \sqrt{\frac{1}{s_0}+\frac{W_w + W_s}{2}}\, .
\label{n0est}
\end{equation}
This equation defines the region which can be explored by any experiment,
specified by its size times running time ($N_0$) and purity factor ($s_0$),
in the sense that a point in the ($\sin^2 2\theta$, $\Delta m^2$) plane
can be studied for seasonal modulations at the $n$-sigma level,
if the corresponding $W_w$ and  $W_s$ satisfy the above constraint.

Notice that even when
the signal to background ratio $s_0$ is quite small, modulations can be
detected if $N_0$ is large enough, as it is obvious since for constant
background statistical fluctuations become less and less important as
the number of counts grows.

In addition, we note that for a high counting rate experiments like Borexino
get even more detailed information by exploring higher moments of the time
dependence and/or resorting to  optimal filtering theory.
The semi-annual variation
we consider is thus to be interpreted as a subset of  the extremely
rich time information which can be exploited by Borexino.
\begin{table}
\caption[aaaa]{
Most recent  experimental data (Experiment), and the
corresponding predictions of the two standard solar models by
Bahcall and Pinsonneault that we use in this work:
BP92~\protect\cite{Bahcall92} and
BP95~\protect\cite{Bahcall95b}.
The errors of the theoretical predictions are 1$\sigma$ effective errors.
\label{tbl1}
              }
\begin{tabular}{lllll}
  & \multicolumn{1}{c}{Chlorine}
  & \multicolumn{1}{c}{Kamiokande}
  & \multicolumn{2}{c}{Gallium \protect[SNU]}                                \\
  & \multicolumn{1}{c}{[SNU]}
  & \multicolumn{1}{c}{ [$10^{6}$ cm$^{-2}$ s$^{-1}$] }
  & \multicolumn{1}{c}{GALLEX}
  & \multicolumn{1}{c}{SAGE}                                                 \\
\tableline
Experiment
&   2.55 $\pm$  0.17 $\pm$ 0.18  \tablenote{Ref.~\protect\cite{Davis}}
&   2.73 $\pm$  0.17 $\pm$ 0.34  \tablenote{Ref.~\protect\cite{Kamio}}
&  77    $\pm$  8    $\pm$ 5     \tablenote{Ref.~\protect\cite{GALLEX}}
&  69    $\pm$ 11    $\pm$ 6     \tablenote{Ref.~\protect\cite{SAGE}}        \\
BP92
&   8.0\phantom{0}  $\pm$  1.0\phantom{0}
&   5.69            $\pm$  0.80
&  \multicolumn{2}{c}{ 132$^{+7}_{-6}$\phantom{00} }                         \\
BP95
&   9.3\phantom{0}  $\pm$  1.1\phantom{0}
&   6.62            $\pm$  1.06
&  \multicolumn{2}{c}{ 137$^{+8}_{-7}$\phantom{00} }                         \\
\end{tabular}
\end{table}
\begin{table}
\caption[bbbb]{
Five choices of parameters within the 90\%~C.L. region of the VO solution:
the ones yielding the best fits if neutrinos are active (first
column) and sterile (second column), the choices that yield the maximum
(third column) and minimum (forth column) beryllium signal and the one
for which seasonal variations are best detectable (last column), i.e.
the one for which $N_0$, see Eq.~(\ref{n0est}), is minimum.
For each choice we show the two oscillation parameters, the
four-experiment $\chi^2$ both for active and sterile neutrinos,
the predictions
for the chlorine, gallium and Kamiokande (active and sterile neutrinos)
experiments, the prediction for the $^7$Be signal
both for a CC+NC detector and for
a CC only detector (or sterile neutrinos) and, in the last row,
$N_0$.
Both Kamiokande and $^7$Be signals are in fractions of the SSM value.
\label{tbl2}
               }
\begin{tabular}{lccccc}
                        & best fit & best fit  & max & min
                        & best  \\
                        & (active) & (sterile) &  $\Phi$(Be)  & $\Phi$(Be)
                        & seasonal    \\
\tableline
  $\sin^{2}2\theta$     & 0.864    & 0.803     & 1.000    & 0.803      & 0.803
  \\
  $\Delta m^{2} ~[10^{-10}$~eV$^{2}]$
                        & 0.615    & 0.625     & 0.577    & 0.645      & 0.901
  \\
$\chi^{2}$  (active)    & 2.4      & 3.0       & 4.0      & 4.4        & 5.9
  \\
$\chi^{2}$  (sterile)   & 7.4      & 6.2       & 11.7     & 8.6        & 15.4
  \\
   Cl $[$SNU$]$         & 2.64     & 2.80      & 2.80     & 2.56       & 2.86
  \\
   Ga $[$SNU$]$         & 66       & 65        & 77       & 60         & 73
  \\
   Ka (active)          & 0.42     & 0.45      & 0.37     & 0.43       & 0.35
  \\
   Ka (sterile)         & 0.32     & 0.36      & 0.26     & 0.36       & 0.23
  \\
   $^{7}$Be (CC+NC)     & 0.59     & 0.50      & 0.98     & 0.38       & 0.63
  \\
   $^{7}$Be (CC only)   & 0.48     & 0.37      & 0.97     & 0.23       & 0.53
  \\
 N$_{0}$ (Borexino)     & 1000     & 1500      & 7$\cdot 10^{4}$  &  8$\cdot
10^{5}$  & 600      \\
\end{tabular}
\end{table}
\begin{table}
\caption[cccc]{
Stability test (only for active neutrinos).
Four solar models derived from BP92 changing the central
temperature ($\pm 2\%$) and the astrophysical factor $S_{17}$ ($\pm 50\%$).
For each choice we show the two oscillation parameters, the
four-experiment $\chi^2$, the predictions
for the chlorine, gallium and Kamiokande
experiments and the prediction for the $^7$Be signal
both for a CC+NC detector and for
a CC only detector.
Both Kamiokande and $^7$Be signals are in fractions of the SSM value.
\label{tbl3}
               }
\begin{tabular}{lcccc}
                        & T$_{c}$/T$^{\text{SSM}}_{c}    = 1.02 $  &
                          T$_{c}$/T$^{\text{SSM}}_{c}    = 0.98 $  &
                          $^{8}$B/$^{8}$B$^{\text{SSM}}  = 1.50 $  &
                          $^{8}$B/$^{8}$B$^{\text{SSM}}  = 0.50 $  \\
\tableline
  $\sin^{2}2\theta$                    & 1.000    & 0.600     & 0.923    &
0.747    \\
  $\Delta m^{2} ~[10^{-10}$~eV$^{2}]$  & 0.687    & 0.635     & 0.676    &
0.375    \\
$\chi^{2}$                             & 1.8      & 3.1       & 2.1      & 4.2
    \\
 Cl $[$SNU$]$                          & 2.56     & 2.70      & 2.66     & 2.76
    \\
 Ga $[$SNU$]$                          & 74       & 73        & 65       & 67
    \\
     Ka                                & 0.39     & 0.37      & 0.47     & 0.37
    \\
   $^{7}$Be  (CC+NC)                   & 0.69     & 0.56      & 0.55     & 0.48
    \\
   $^{7}$Be  (CC)                      & 0.61     & 0.45      & 0.44     & 0.35
    \\
\end{tabular}
\end{table}
\begin{table}
\caption[dddd]{
Six choices of parameters within the 90\%~C.L. region of the MSW solution:
the first three are within the part of this region at small mixing angle,
while the last three are within the part at large mixing angle
(see Fig.~\protect\ref{fig14}). For each of the two parts, we give the
choice of parameters that yields the best fit (overall best at small angle
in 1st column, local best at large angle in 4th column) and the
choice that yields
the maximum (2nd and 5th column) and minimum (3nd and 6th column)
beryllium signal. All results are for active neutrinos.
We report the two oscillation parameters, the
four-experiment $\chi^2$, the predictions
for the chlorine, gallium and Kamiokande
experiments and the prediction for the $^7$Be signal
for a CC+NC detector.
Both Kamiokande and $^7$Be signals are in fractions of the SSM value.
\label{tbl4}
               }
\begin{tabular}{lcccccc}
                        & \multicolumn{3}{c}{small $\theta$ }
                        & \multicolumn{3}{c}{large $\theta$ }  \\
                        & best fit & $\Phi_{\text{max}}$(Be) &
$\Phi_{\text{min}}$(Be)
                        & best fit & $\Phi_{\text{max}}$(Be) &
$\Phi_{\text{min}}$(Be) \\
\tableline
$\sin^{2}2\theta$       & $5.8   \cdot 10^{-3}$  & $4.2   \cdot 10^{-3}$  &
$7.9   \cdot 10^{-3}$
                        & 0.732                  & 0.626                  &
0.732                    \\
$\Delta m^{2} ~[10^{-5}$~eV$^{2}]$
                        & 0.624                  & 1.052                  &
0.415
                        & 2.990                  & 6.000                  &
0.992                    \\
$\chi^{2}$  (active)    & 0.5                    & 4.7                    & 4.7
                        & 2.7                    & 5.0                    & 4.9
                     \\
 Cl $[$SNU$]$           & 2.50                   & 2.53                   &
2.64
                        & 2.70                   & 2.73                   &
2.54                     \\
 Ga $[$SNU$]$           & 75                     & 89                     & 60
                        & 72                     & 81                     & 60
                     \\
 Ka                     & 0.51                   & 0.45                   &
0.53
                        & 0.39                   & 0.36                   &
0.41                     \\
 $^{7}$Be  (CC+NC)      & 0.22                   & 0.50                   &
0.20
                        & 0.64                   & 0.71                   &
0.50                     \\
 $^{7}$Be (CC only)     & 0.03                   & 0.37                   &
0.01
                        & 0.54                   & 0.64                   &
0.37                     \\
\end{tabular}
\end{table}
%
%
%
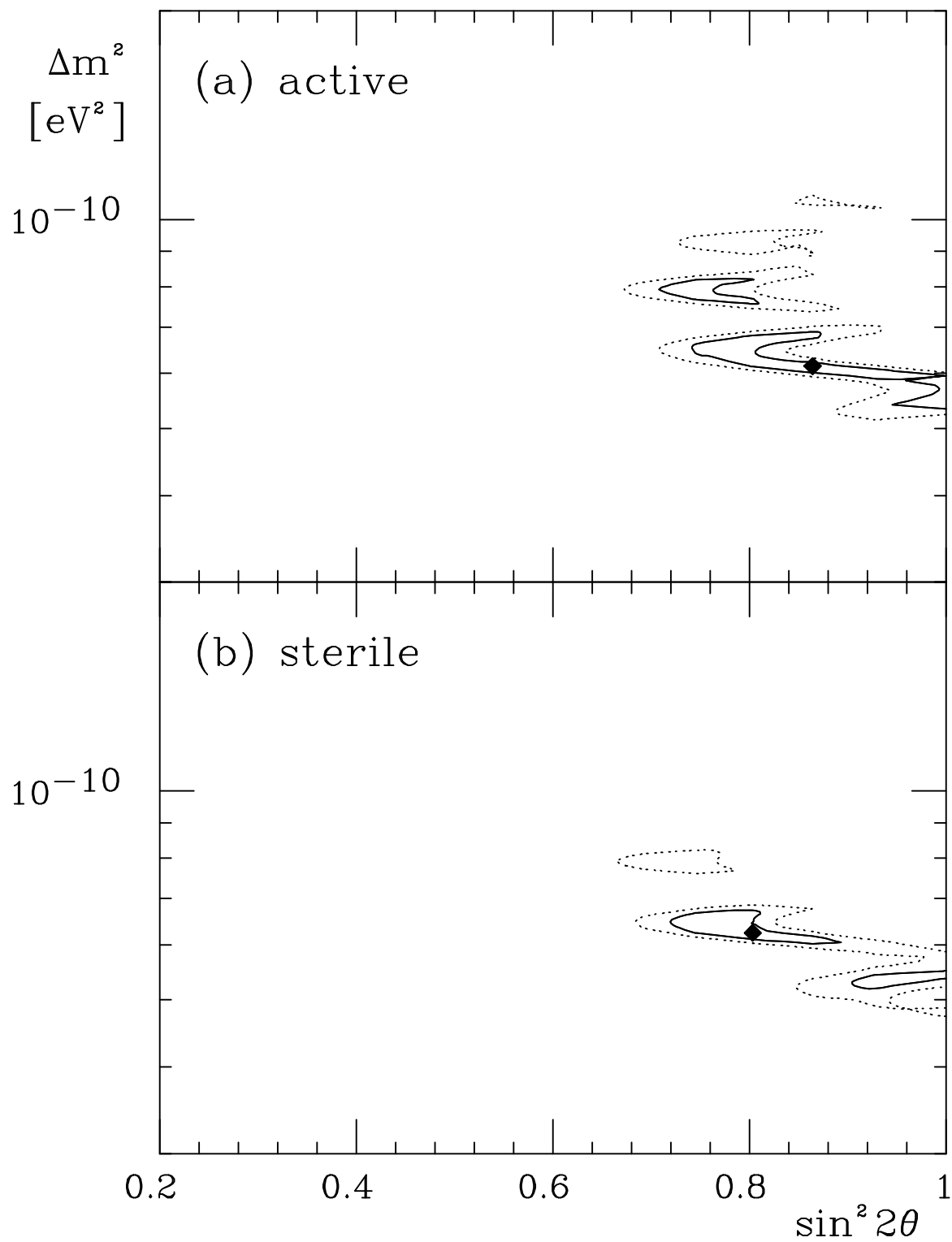
\begin{figure}
\caption[clrga]{
Regions of parameters allowed at the 68\%
(solid line, $\chi^2< \chi^2_{\text{min}} + 2.28$)
and 95\% (dots, $\chi^2 < \chi^2_{\text{min}} + 5.99$)
C.L. for oscillations
into (a) active, and (b) sterile neutrinos obtained combining data from
all 4 experiments (see Table~\protect\ref{tbl1}).
We have used fluxes from the SSM of
Bahcall and Pinsonneault with helium diffusion~\protect\cite{Bahcall92},
and no theoretical errors.
Information about the best fit points (diamond)
and several other relevant points can be found in Table~\protect\ref{tbl2}.
        }
\label{fig1}
\end{figure}
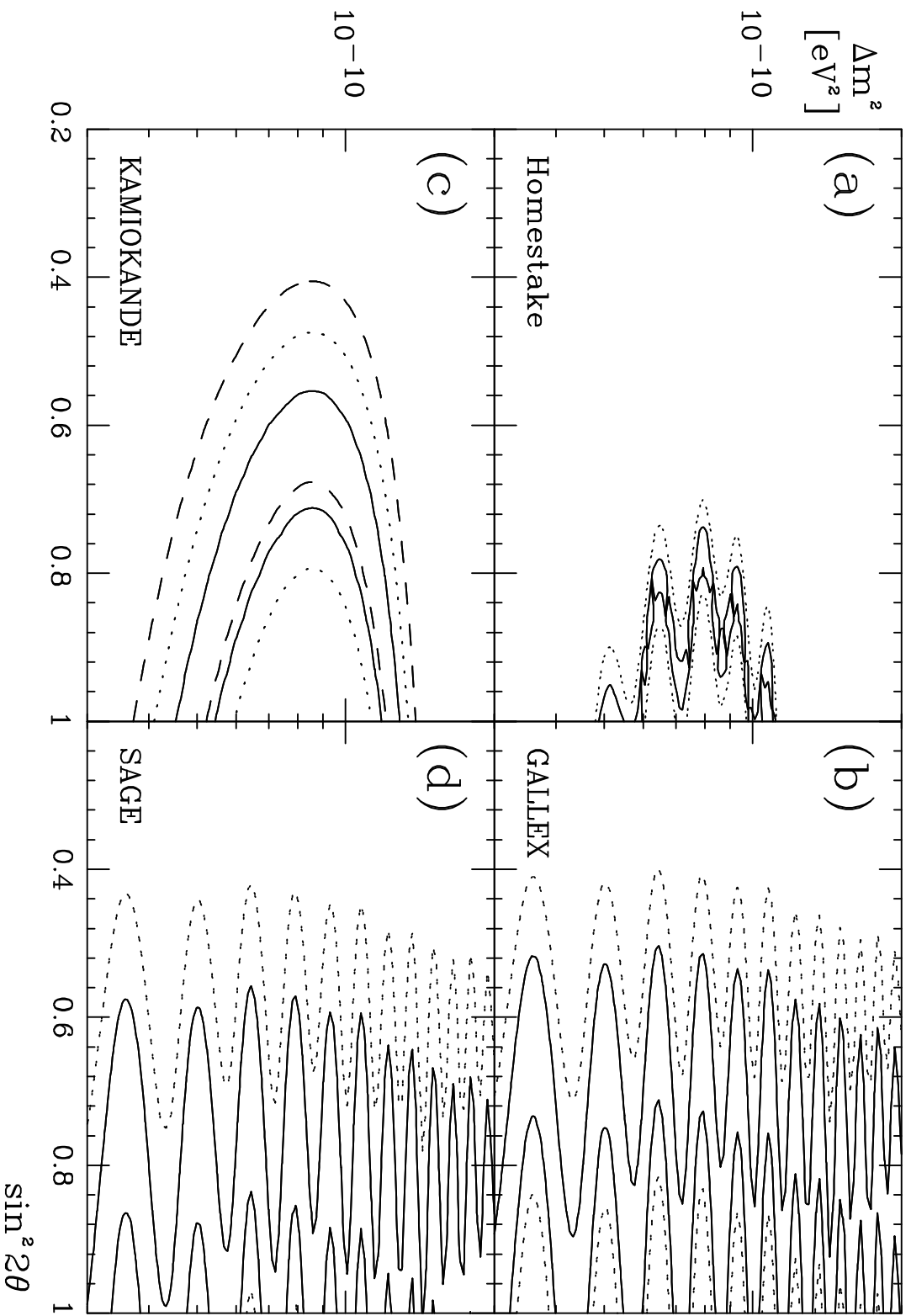
\begin{figure}
\caption[clrgs]{
Regions of parameters allowed at the 68\% (solid line, $\chi^2< 0.99$)
and 95\% (dots, $\chi^2< 3.84$)
C.L. by each of the 4 experiments separately: (a) chlorine, (b) GALLEX,
(c) Kamiokande and (d) SAGE.  For the Kamiokande (d), we show both
confidence regions for the case of active neutrinos, and only the 95\%
C.L. region for the case of sterile neutrinos (dashes).
Fluxes as in Fig.~\protect\ref{fig1}.
        }
\label{fig2}
\end{figure}
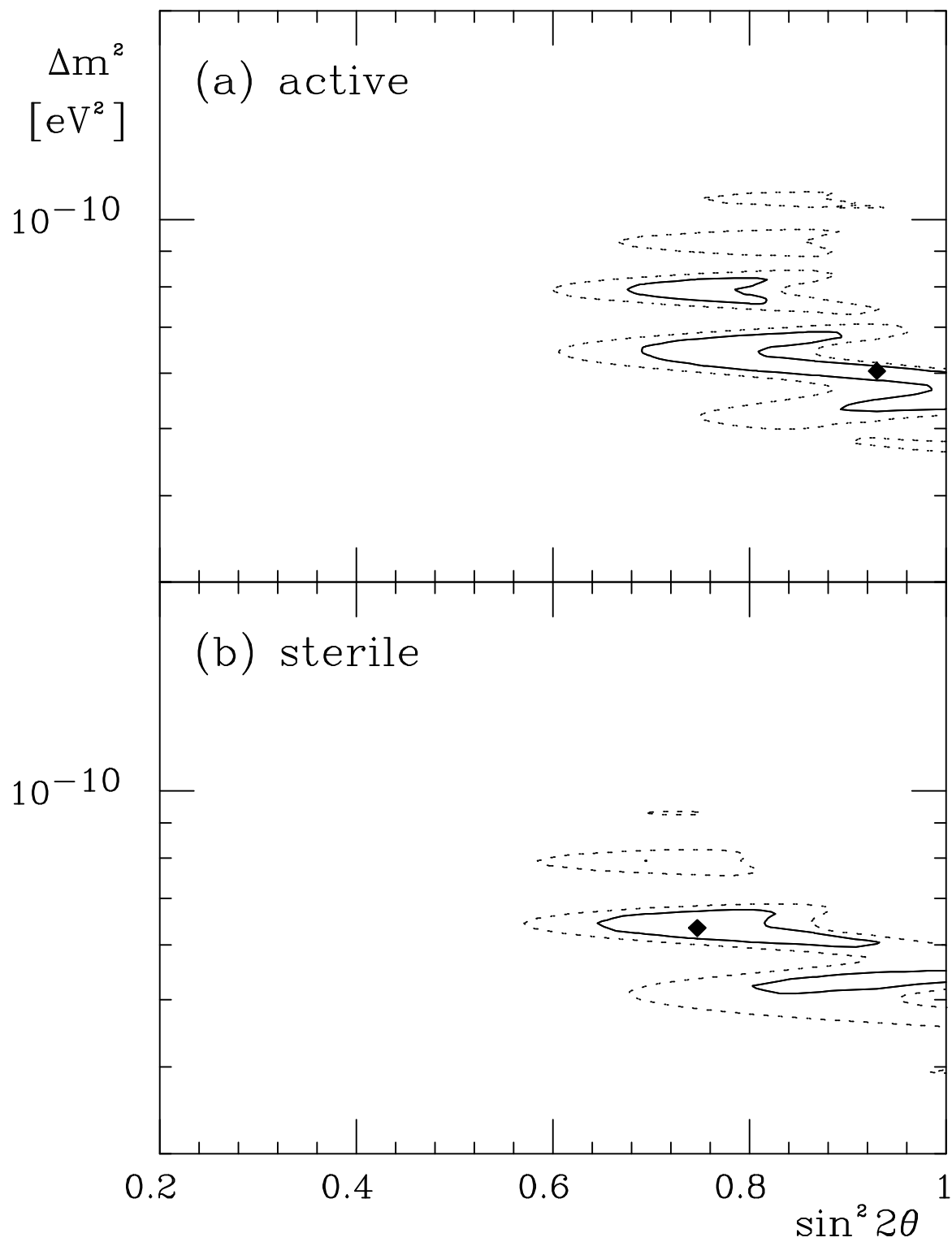
\begin{figure}
\caption[clrgth]{
Same as Figure~\protect\ref{fig1} but taking into account the
properly correlated theoretical uncertainties on the neutrino fluxes.
        }
\label{fig3}
\end{figure}
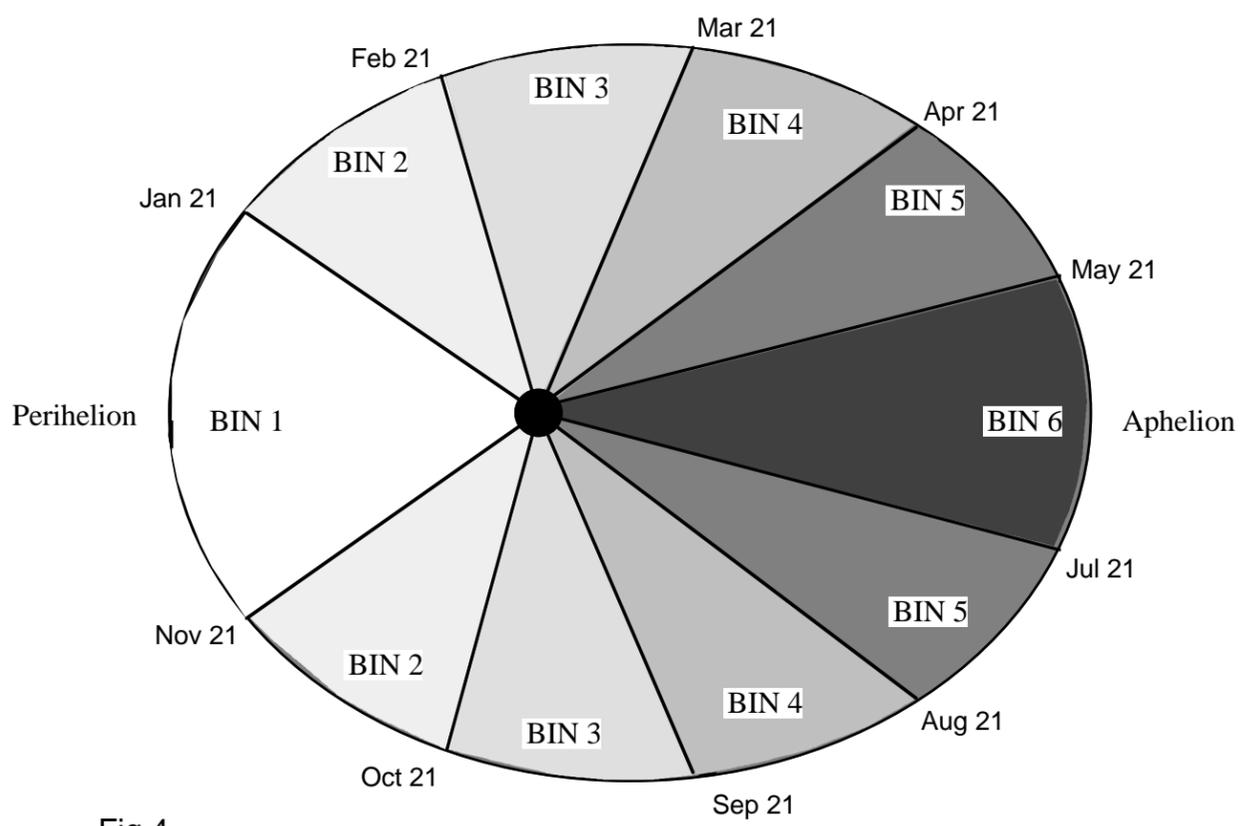
\begin{figure}
\caption[bins]{
The six periods (bins) into which the year has been divided. Bins contain
equal areas.
        }
\label{fig4}
\end{figure}
\begin{figure}
\caption[tvar]{Signals from the (a) chlorine~\protect\cite{Davis2},
(b) GALLEX~\protect\cite{GALLEX2}, (c) Kamiokande~\protect\cite{Kamio2}
and (d) SAGE~\protect\cite{SAGE2}
experiments as functions of the
absolute time difference from the winter solstice (perihelion) in
fractions of year (0.5 is then the aphelion).
Bins as defined in Fig.~\protect\ref{fig4} have been used.
The solid line is the
theoretical prediction for the best fit parameters (active neutrinos).
The dashed line
is the case of maximal $^7$Be flux within the 90\% C.L. region,
while the dotted line gives the maximal seasonal variation within the
same region. These three lines correspond to parameters in columns
1, 3 and 5 of Table~\protect\ref{tbl2}.
Signals are normalized to the yearly averages.
        }
\label{fig5}
\end{figure}
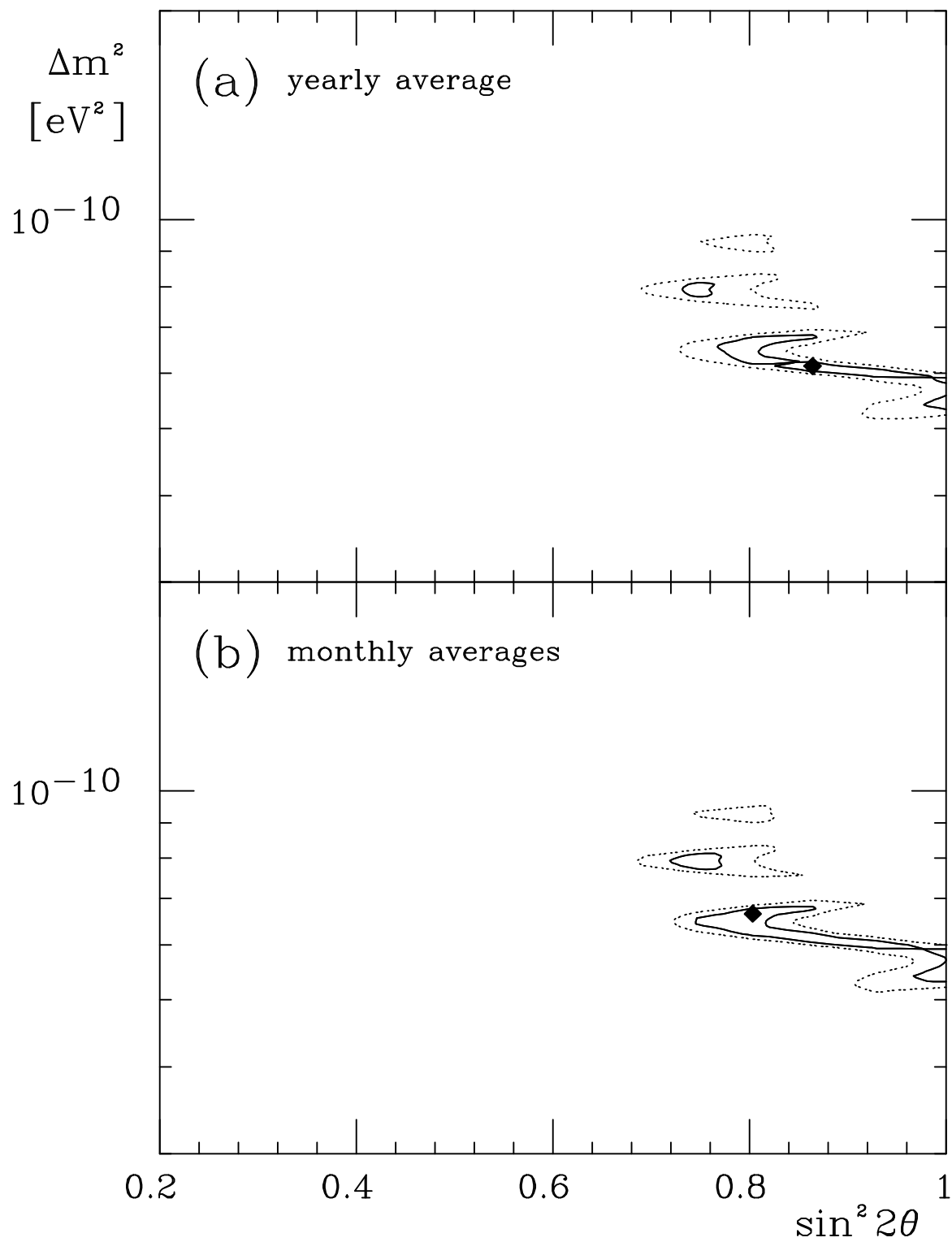
\begin{figure}
\caption[clrgtv]{
Confidence regions allowed at the 68\% (solid line,
$\chi^2< \chi^2_{\text{min}} + 2.28$)
and 95\% (dots, $\chi^2< \chi^2_{\text{min}} + 5.99$)C.L. for oscillations
into active neutrinos obtained comparing the binned experimental data to
(a) constant (averaged over a year), and (b) time-dependent
(averaged over a month) theoretical predictions.
Fluxes are from the SSM of Bahcall and Pinsonneault with
helium diffusion~\protect\cite{Bahcall92},
and no theoretical errors.
        }
\label{fig6}
\end{figure}
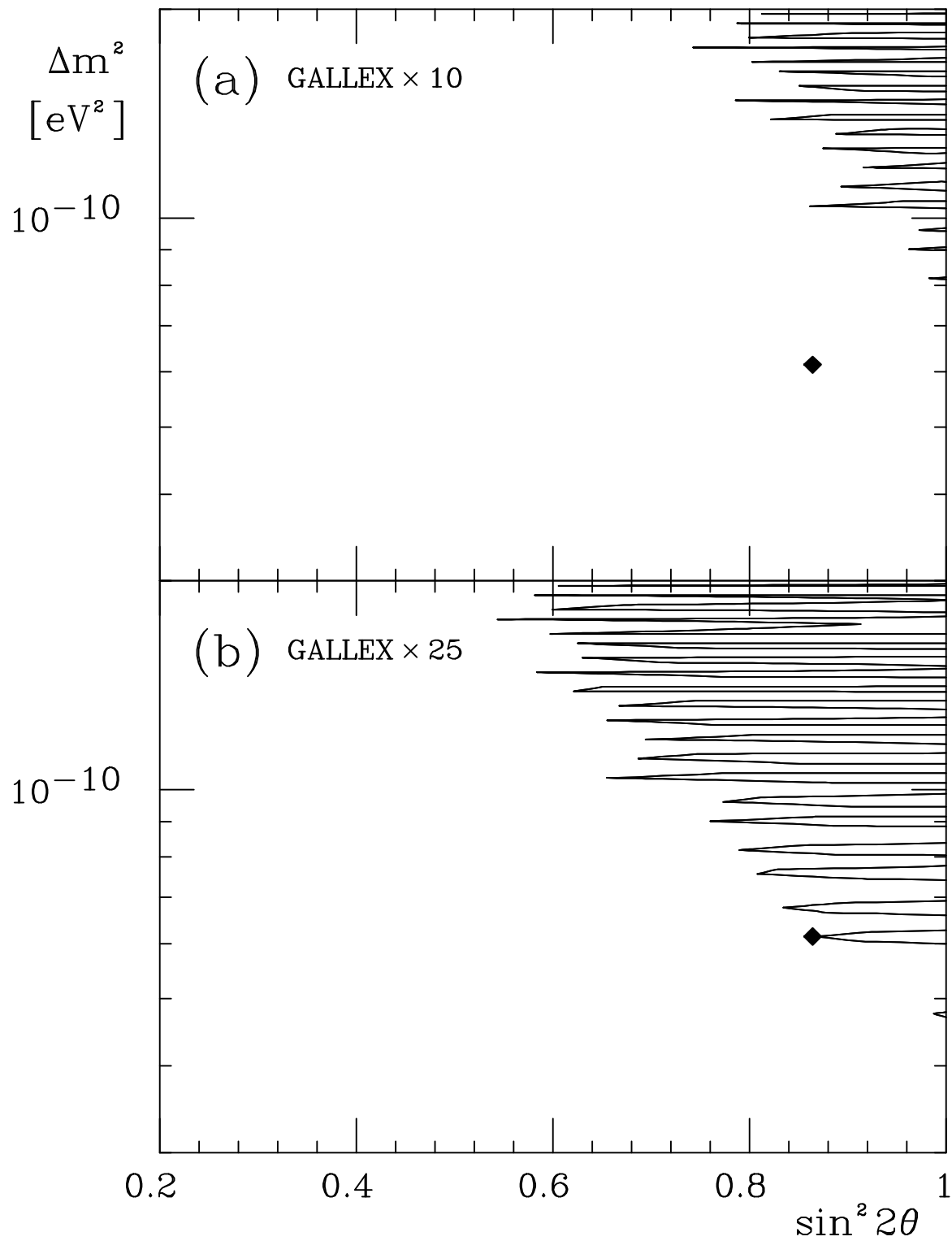
\begin{figure}
\caption[svga]{
Regions of the parameter space that gallium experiments can explore
at the three sigma level by just looking for semi-annual variations
if they are able to collect a statistics
(a) 10 times and (b) 25 times the present GALLEX statistics (30 tons
$\times$ 4 years). For reference we have also reported the best fit point
from Fig.~\protect\ref{fig1}a.
        }
\label{fig7}
\end{figure}
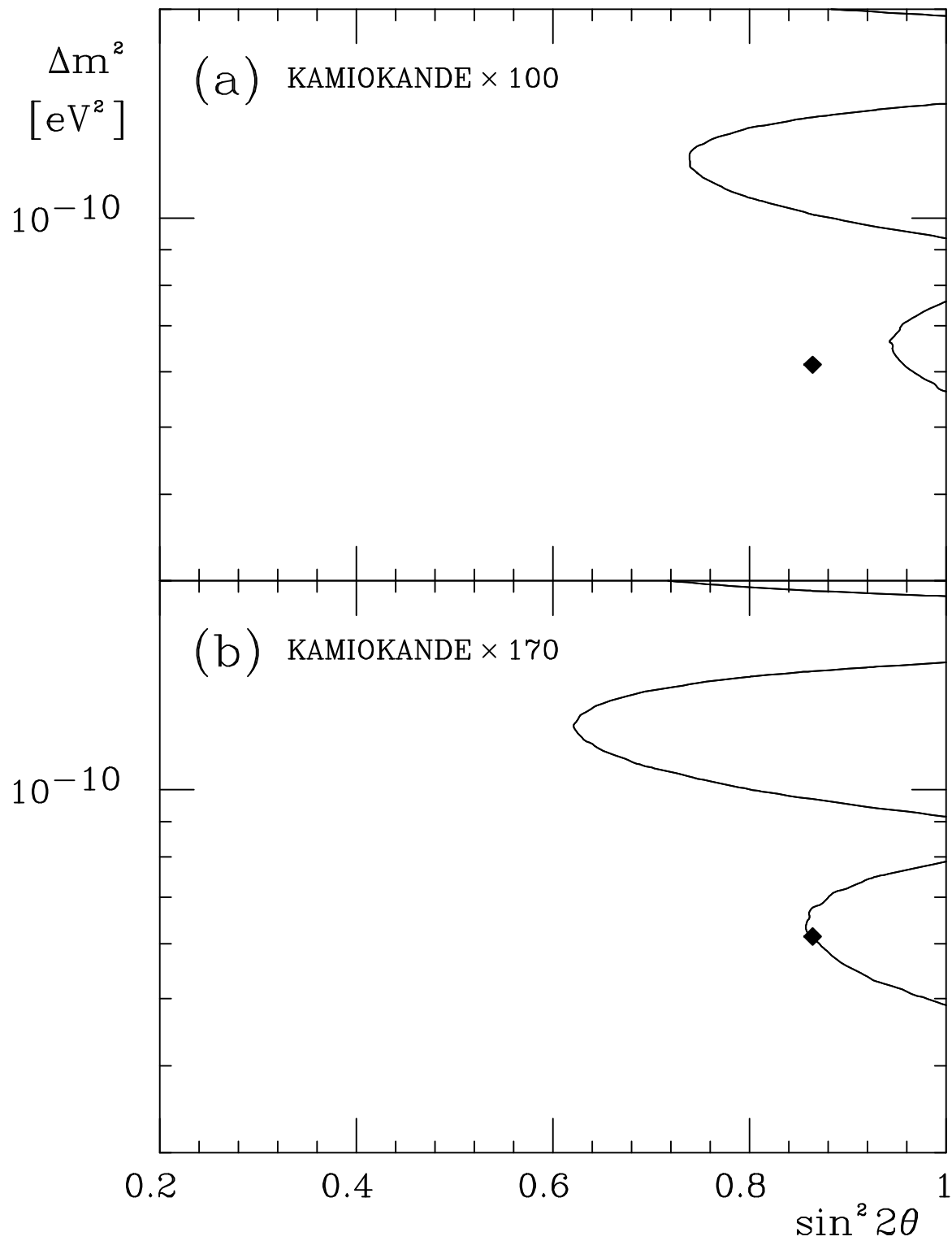
\begin{figure}
\caption[svSK]{
Regions of the parameter space that SuperKamiokande should be able
to explore at the three sigma level by just looking for seasonal variations
once it has collected
a statistics about (a) 100 times and (b) 170 times the present
Kamiokande statistics, if neutrinos are active; regions are obviously
larger if neutrinos are sterile.
For reference we have also reported the best fit point
from Fig.~\protect\ref{fig1}a.
        }
\label{fig8}
\end{figure}
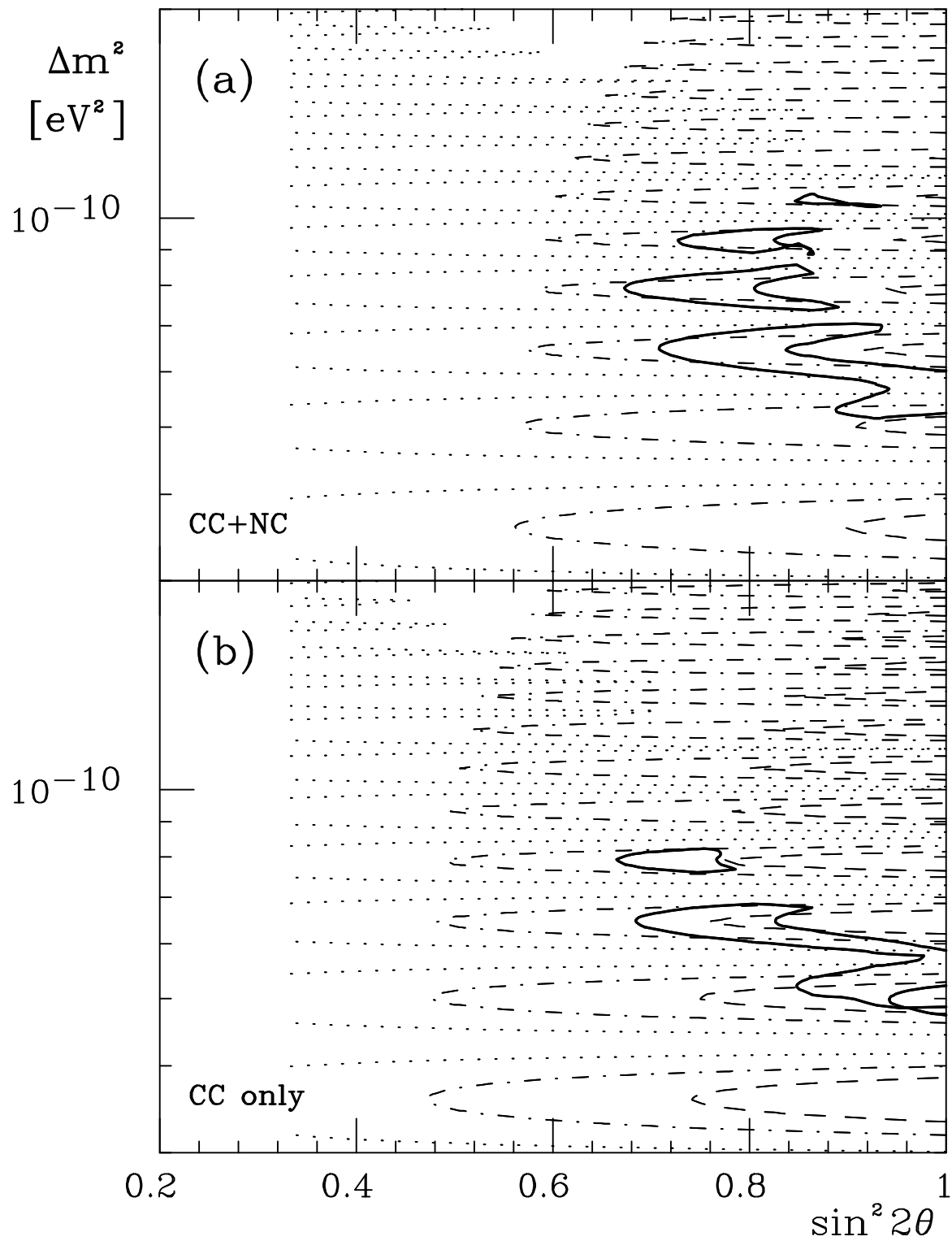
\begin{figure}
\caption[isobe]{
Curves of iso-suppression of the $^7$Be neutrino signal:
(a) active neutrinos and CC + NC detector, and (b)
CC only detector or sterile neutrinos. We show only three iso-curves
0.3 (dashes), 0.6 (dashes and dots) and 0.9 (dots).
Curves bounding the 95\% confidence regions (solid line,
see Fig.~\protect\ref{fig1}) are superposed to the curves of
iso-suppression.
        }
\label{fig9}
\end{figure}
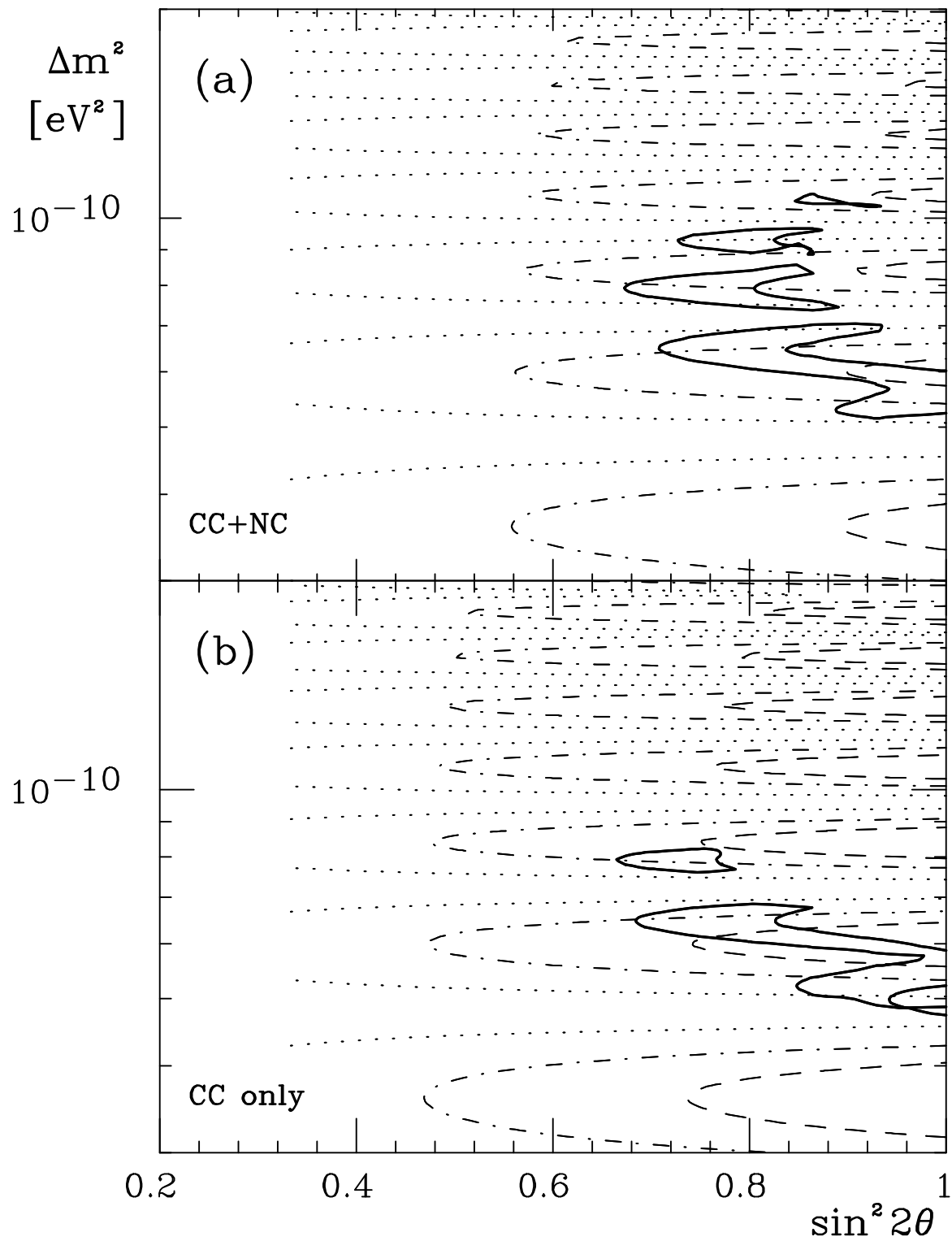
\begin{figure}
\caption[isopep]{
Same as Fig.~\protect\ref{fig9} for pep neutrinos.
        }
\label{fig10}
\end{figure}
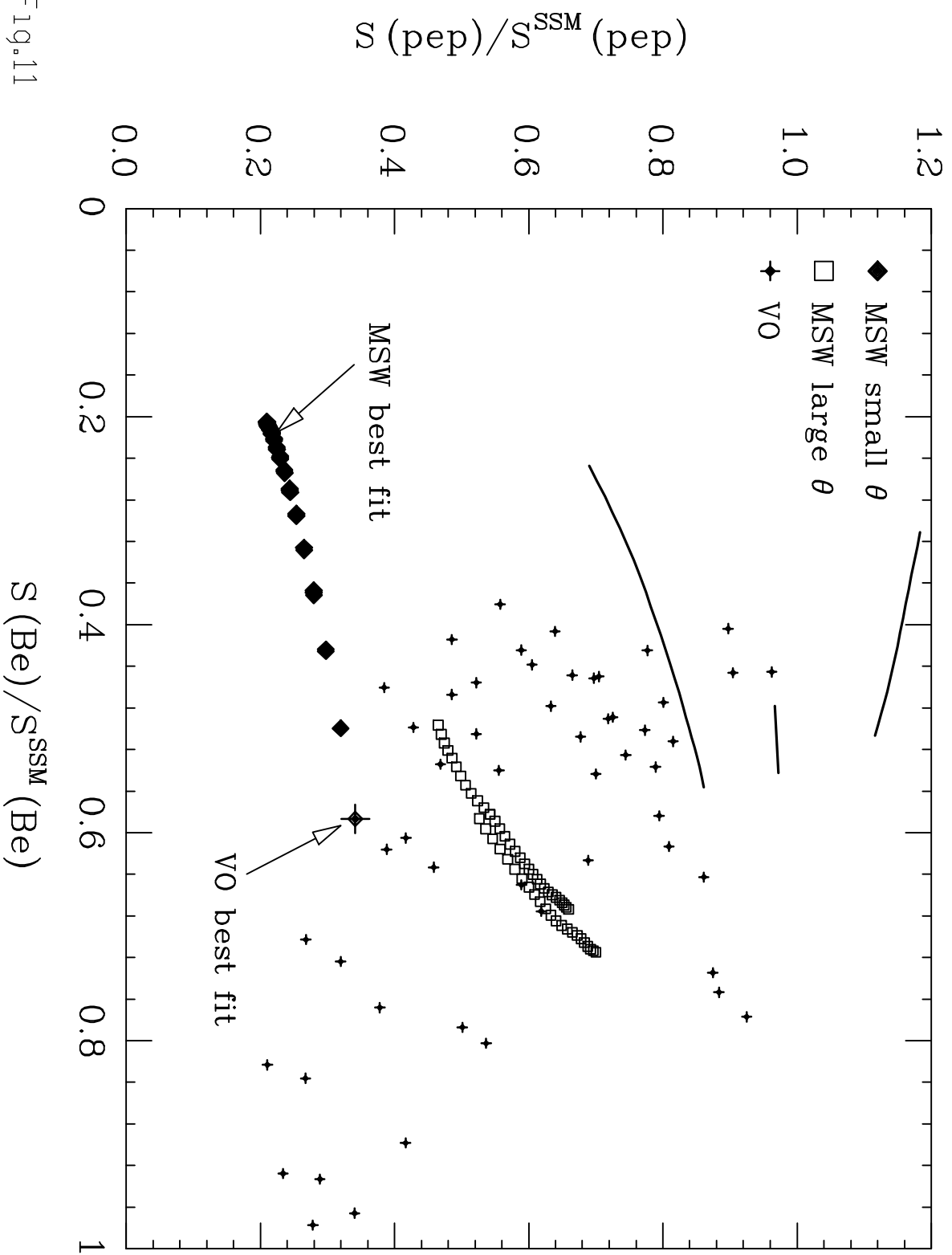
\begin{figure}
\caption[bepep]{The pep neutrino signal vs. the $^7$Be
neutrino signal as predicted by different models for active neutrinos and
CC + NC detection. Signals are in units of the corresponding BP92 fluxes.
The three solid lines in the upper part give the range of predictions for
several non standard solar models (see Ref.~\protect\cite{Cast94b}).
Crosses are VO predictions from points uniformly distributed
inside the  90\% C.L. region; the best fit point is bigger and
explicitly labeled.
Filled diamonds
are MSW predictions for parameters uniformly distributed inside the
the small-mixing-angle part of the 90\% C.L. region; the
best fit point is explicitly labeled. Open squared ($\Box$)
are MSW predictions from points uniformly distributed
inside the large-mixing-angle part of the 90\% C.L. region.
        }
\label{fig11}
\end{figure}
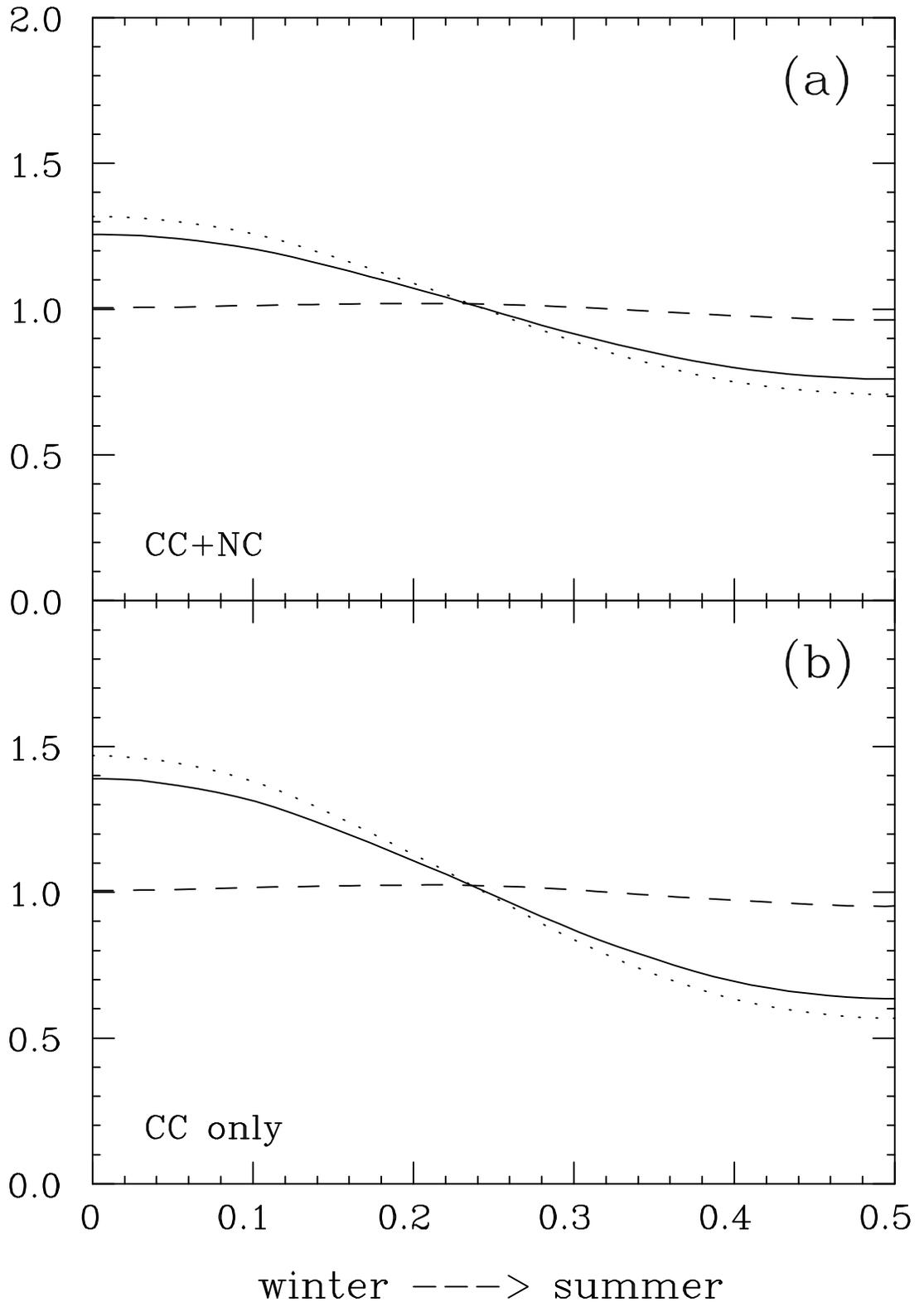
\begin{figure}
\caption[bevar]{
Expected variation of the $^7$Be signal as a function of the
absolute time difference from the perihelion in fractions of year
(0.5 is then the aphelion).
The solid line is the
theoretical prediction for the best fit parameters. The dashed line
is the case of maximal $^7$Be flux within the 90\% C.L. region,
while the dotted line gives the maximal seasonal variation within the
same region. These three lines correspond to parameters in column
1, 3 and 5 of Table~\protect\ref{tbl2}.
Signals are normalized so that the yearly average is one.
We show the signal both for (a) CC + NC detector and active neutrinos
and for (b) CC only detector or sterile neutrinos;
NC cross section has been assumed
1/5 of the CC cross section.
Variations due to the change of the solid angle are not included.
        }
\label{fig12}
\end{figure}
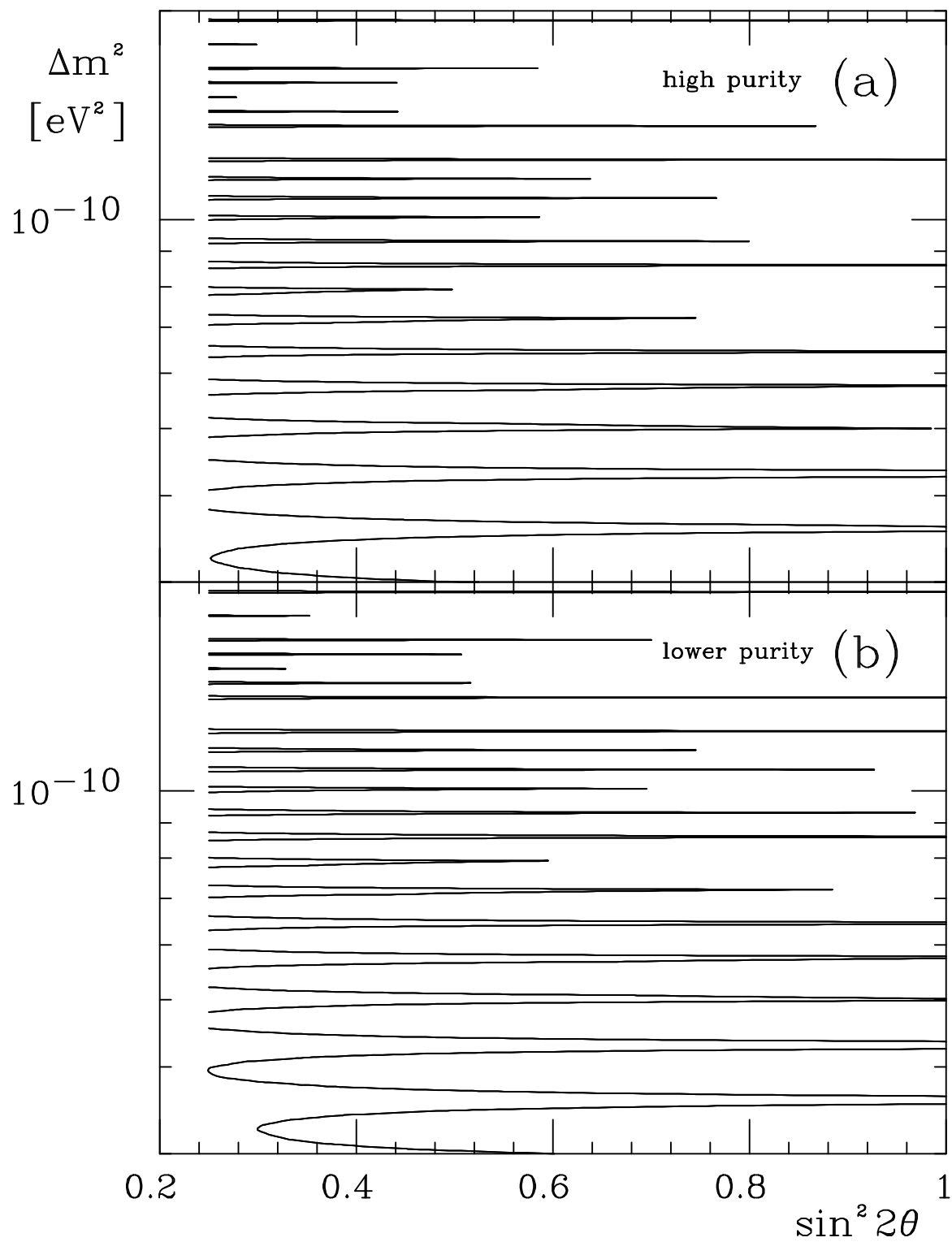
\begin{figure}
\caption[n0]{
Regions of the parameter space that Borexino should be able
to explore at the three sigma level by just looking for seasonal variation
once it has run for three years, which correspond to about fifty thousand
events for the SSM beryllium flux
(50 events per day times 1000 days). We consider
two possible backgrounds:
(a) 11 background events per day, which corresponds to
the expected high purity for the final full-size experiment, and
(b) 33 background events per day, which corresponds to
about the purity that has already been obtained~\cite{bonetti}.
        }
\label{fig13}
\end{figure}
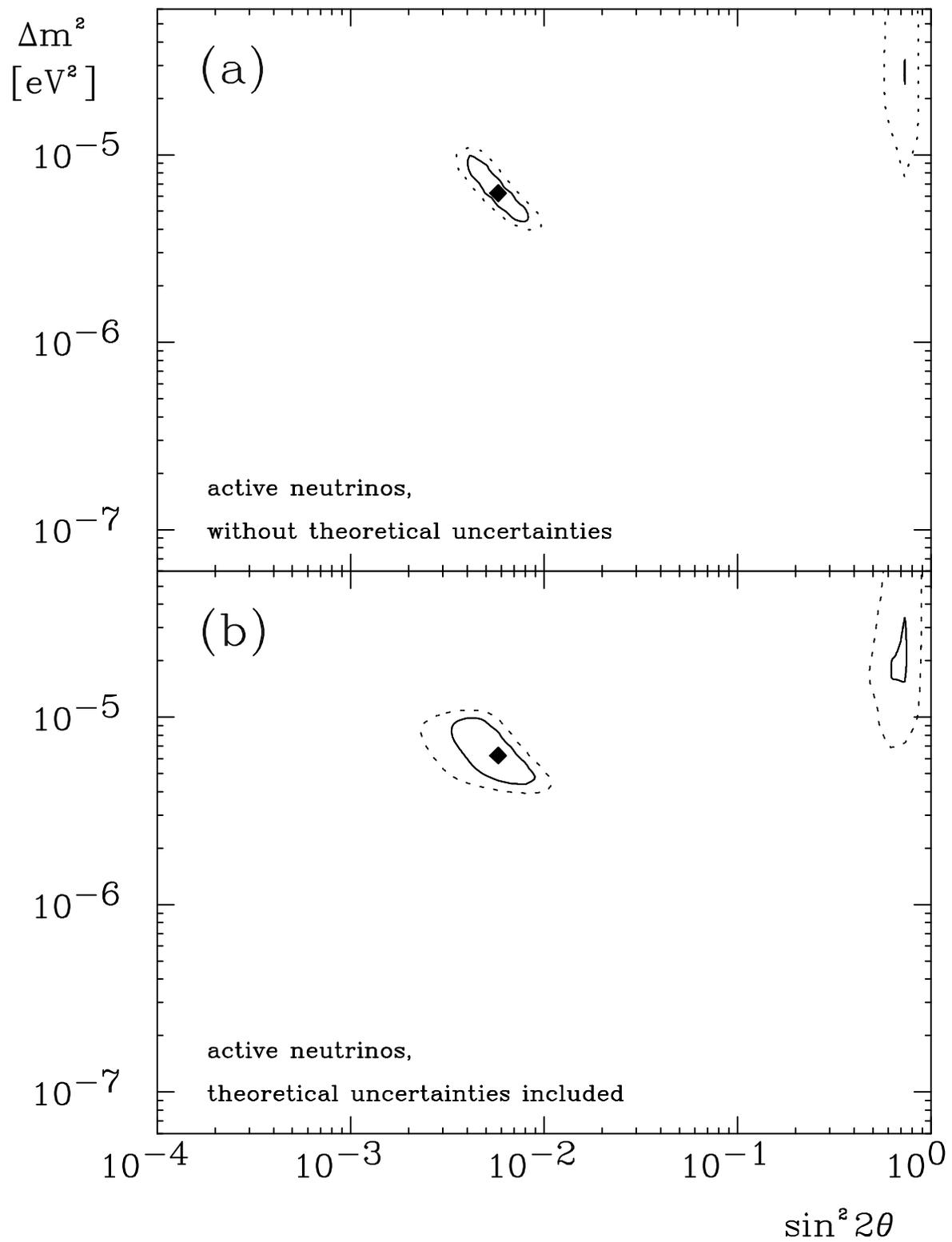
\begin{figure}
\caption[msw]{
MSW confidence regions using the model by Bahcall and Pinsonneault (a) not
including and (b) including properly correlated theoretical errors.
Solid curves delimit 68\% CL regions, dashed curves
95\% CL regions. The plotted point marks the best fit. Calculation as
in Ref.~\protect\cite{Fiore}, but with the new experimental data.
Mixing is with active neutrinos.
        }
\label{fig14}
\end{figure}
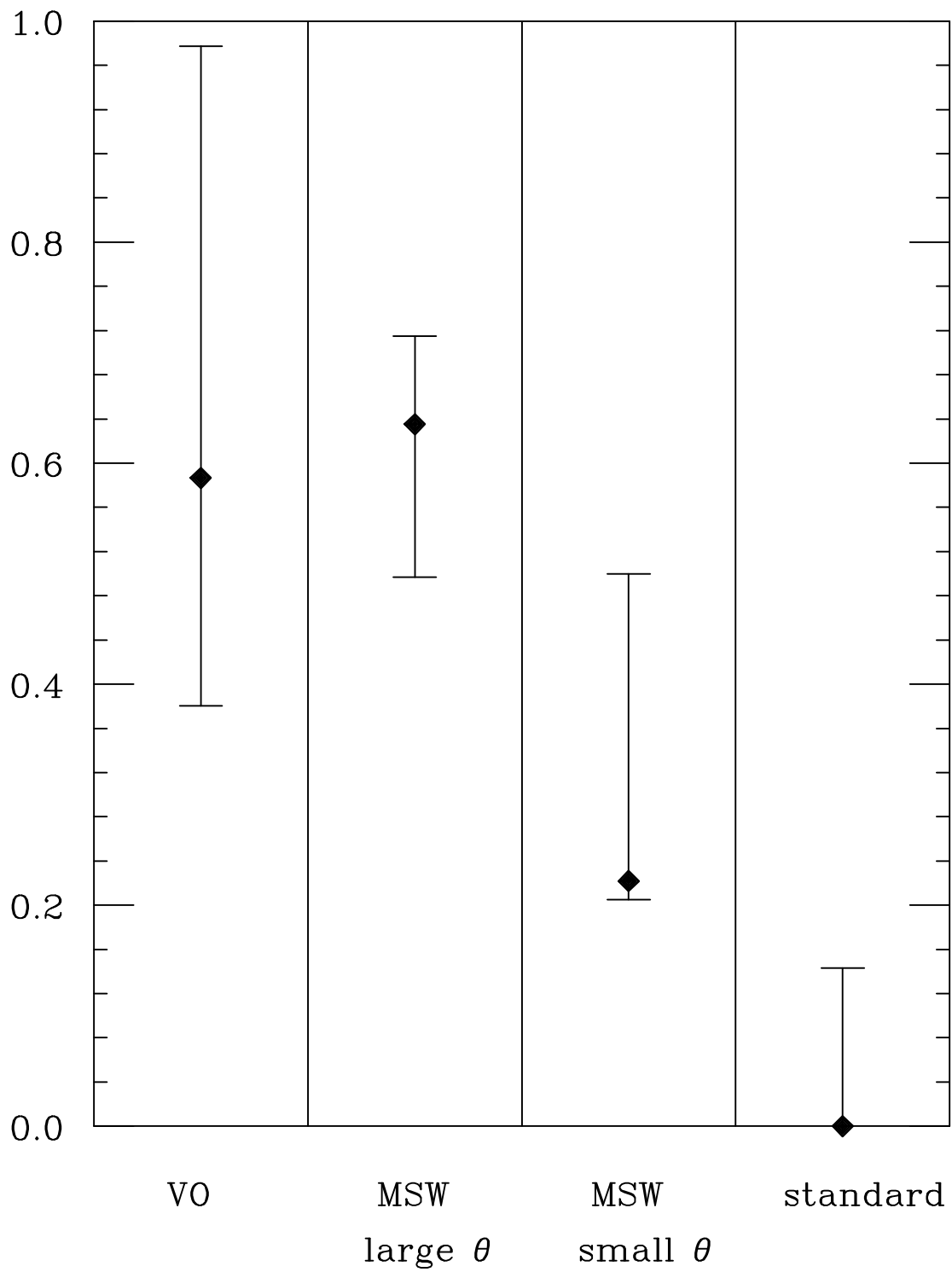
\begin{figure}
\caption[new]{
Suppressions of the $^7$Be signal for a CC + NC detector (e.g. Borexino)
relative to its SSM value. We show predictions in the case of mixing with
active neutrinos for vacuum oscillations (VO), large-mixing-angle MSW
(large $\theta$), small-mixing-angle MSW (small $\theta$), and
standard neutrinos (standard). The intervals indicate the ranges of values
allowed by each solution at the 90\% C.L.,
while the best fit values within each region are marked by filled diamonds.
        }
\label{fig15}
\end{figure}
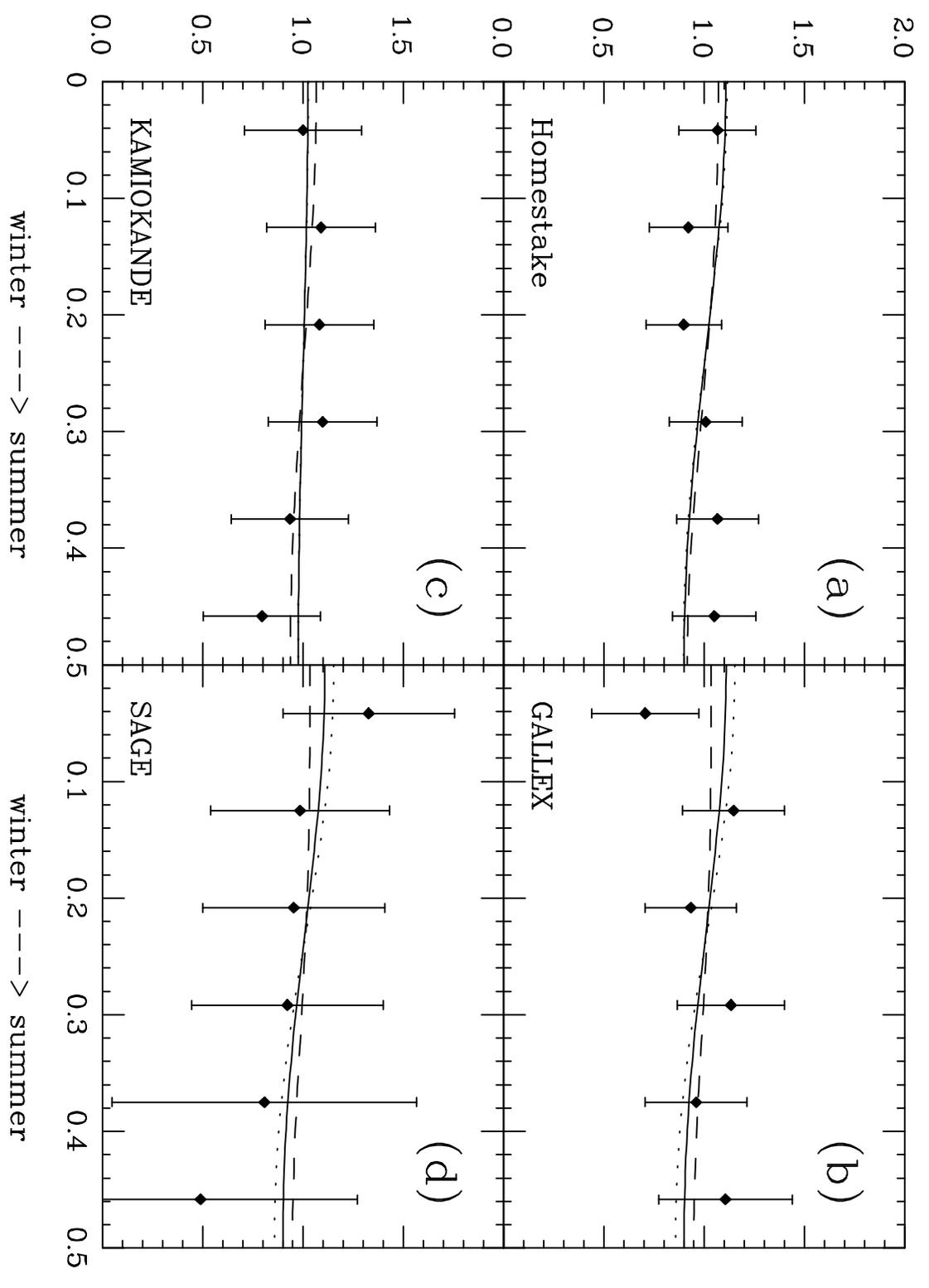
\end{document}